\documentclass[twocolumn]{aastex631}

\usepackage{gensymb}
\usepackage{amsmath}
\usepackage{tikz}
\usepackage{subfigure}
\usepackage{xurl}

\newcommand{\annotatedimage}[2]{%
\begin{tikzpicture}
    \node[anchor=south west, inner sep=0] (img) at (0,0)
        {\includegraphics[width=\columnwidth]{#1}};
    \node[anchor=north west, font=\bfseries] at (img.north west) {#2};
\end{tikzpicture}%
}

\shorttitle{South Pole Atmosphere for CMB}
\shortauthors{BICEP/\textit{Keck} Collaboration}

\begin{document}

\title{BICEP/\textit{Keck} XXII: Analysis of the South Pole Atmosphere for CMB Observations}

\author{BICEP/\textit{Keck} Collaboration: P. A. R. Ade}
\affiliation{School of Physics and Astronomy, Cardiff University, Cardiff, CF24 3AA, United Kingdom}
\author{Z. Ahmed}
\affiliation{Kavli Institute for Particle Astrophysics and Cosmology, Stanford University, Stanford, CA 94305, USA}
\affiliation{SLAC National Accelerator Laboratory, Menlo Park, CA 94025, USA}
\author{M. Amiri}
\affiliation{Department of Physics and Astronomy, University of British Columbia, Vancouver, British Columbia, V6T 1Z1, Canada}
\author{D. Barkats}
\affiliation{Center for Astrophysics, Harvard \& Smithsonian, Cambridge, MA 02138, USA}
\author{D. Barron}
\affiliation{Department of Physics and Astronomy, University of New Mexico, Albuquerque, NM 87131, USA}
\author{R. Basu Thakur}
\affiliation{Department of Physics, California Institute of Technology, Pasadena, CA 91125, USA}
\author{I. Birdwell}
\affiliation{Department of Physics and Astronomy, University of New Mexico, Albuquerque, NM 87131, USA}
\author{C. A. Bischoff}
\affiliation{Department of Physics, University of Cincinnati, Cincinnati, OH 45221, USA}
\author{D. Beck}
\affiliation{Department of Physics, Stanford University, Stanford, CA 94305, USA}
\author{J. J. Bock}
\affiliation{Department of Physics, California Institute of Technology, Pasadena, CA 91125, USA}
\affiliation{Jet Propulsion Laboratory, California Institute of Technology, Pasadena, CA 91109, USA}
\author{V. Buza}
\affiliation{Kavli Institute for Cosmological Physics, University of Chicago, Chicago, IL 60637, USA}
\author{B. Cantrall}
\affiliation{Department of Physics, Stanford University, Stanford, CA 94305, USA}
\affiliation{Kavli Institute for Particle Astrophysics and Cosmology, Stanford University, Stanford, CA 94305, USA}
\author{J. R. Cheshire IV}
\affiliation{Department of Physics, California Institute of Technology, Pasadena, CA 91125, USA}
\author{J. Connors}
\affiliation{National Institute of Standards and Technology, Boulder, CO 80305, USA}
\author{J. Cornelison}
\affiliation{Argonne National Laboratory, High Energy Physics Division, Lemont, IL 60439, USA}
\author{M. Crumrine}
\affiliation{School of Physics and Astronomy, University of Minnesota, Minneapolis, MN 55455, USA}
\author{A. J. Cukierman}
\affiliation{Department of Physics, California Institute of Technology, Pasadena, CA 91125, USA}
\author{E. Denison}
\affiliation{National Institute of Standards and Technology, Boulder, CO 80305, USA}
\author{L. Duband}
\affiliation{Service des Basses Temperatures, Commissariat a l'Energie Atomique, 38054 Grenoble, France}
\author{M. A. Echter}
\affiliation{Center for Astrophysics, Harvard \& Smithsonian, Cambridge, MA 02138, USA}
\author{M. Eiben}
\affiliation{Faculty of Physical Sciences, University of Iceland, 102 Reykjav\'ik, Iceland}
\author{B. D. Elwood}
\affiliation{Center for Astrophysics, Harvard \& Smithsonian, Cambridge, MA 02138, USA}
\affiliation{Department of Physics, Harvard University, Cambridge, MA 02138, USA}
\author{S. Fatigoni}
\affiliation{Department of Physics, California Institute of Technology, Pasadena, CA 91125, USA}
\email{sofiaf@caltech.edu}
\author{J. P. Filippini}
\affiliation{Department of Physics, Grainger College of Engineering, University of Illinois Urbana-Champaign, Urbana, IL 61801, USA}
\author{A. Fortes}
\affiliation{Department of Physics, Stanford University, Stanford, CA 94305, USA}
\author{M. Gao}
\affiliation{Department of Physics, California Institute of Technology, Pasadena, CA 91125, USA}
\author{C. Giannakopoulos}
\affiliation{Department of Physics, University of Cincinnati, Cincinnati, OH 45221, USA}
\author{N. Goeckner-Wald}
\affiliation{Department of Physics, Stanford University, Stanford, CA 94305, USA}
\author{D. C. Goldfinger}
\affiliation{Department of Physics, Stanford University, Stanford, CA 94305, USA}
\author{S. Gratton}
\affiliation{Centre for Theoretical Cosmology, Department of Applied Mathematics and Theoretical Physics, University of Cambridge, Cambridge CB3 0WA, United Kingdom}
\affiliation{Kavli Institute for Cosmology, University of Cambridge, Cambridge CB3 0HA, United Kingdom}
\author{J. A. Grayson}
\affiliation{Department of Physics, Stanford University, Stanford, CA 94305, USA}
\author{A. Greathouse}
\affiliation{Department of Physics, California Institute of Technology, Pasadena, CA 91125, USA}
\author{P. K. Grimes}
\affiliation{Center for Astrophysics, Harvard \& Smithsonian, Cambridge, MA 02138, USA}
\author{G. Halal}
\affiliation{Department of Physics, Stanford University, Stanford, CA 94305, USA}
\author{M. Halpern}
\affiliation{Department of Physics and Astronomy, University of British Columbia, Vancouver, British Columbia, V6T 1Z1, Canada}
\author{E. Hand}
\affiliation{Department of Physics, University of Cincinnati, Cincinnati, OH 45221, USA}
\author{S. A. Harrison}
\affiliation{Center for Astrophysics, Harvard \& Smithsonian, Cambridge, MA 02138, USA}
\author{S. Henderson}
\affiliation{Kavli Institute for Particle Astrophysics and Cosmology, Stanford University, Stanford, CA 94305, USA}
\affiliation{SLAC National Accelerator Laboratory, Menlo Park, CA 94025, USA}
\author{T. D. Hoang}
\affiliation{School of Physics and Astronomy, University of Minnesota, Minneapolis, MN 55455, USA}
\author{J. Hubmayr}
\affiliation{National Institute of Standards and Technology, Boulder, CO 80305, USA}
\author{H. Hui}
\affiliation{Department of Physics, California Institute of Technology, Pasadena, CA 91125, USA}
\author{K. D. Irwin}
\affiliation{Department of Physics, Stanford University, Stanford, CA 94305, USA}
\author{J. H. Kang}
\affiliation{Department of Physics, California Institute of Technology, Pasadena, CA 91125, USA}
\author{K. S. Karkare}
\affiliation{Department of Physics, Boston University, Boston, MA 02215, USA}
\author{S. Kefeli}
\affiliation{Department of Physics, California Institute of Technology, Pasadena, CA 91125, USA}
\author{J. M. Kovac}
\affiliation{Center for Astrophysics, Harvard \& Smithsonian, Cambridge, MA 02138, USA}
\affiliation{Department of Physics, Harvard University, Cambridge, MA 02138, USA}
\author{C. Kuo}
\affiliation{Department of Physics, Stanford University, Stanford, CA 94305, USA}
\author{K. Lasko}
\affiliation{School of Physics and Astronomy, University of Minnesota, Minneapolis, MN 55455, USA}
\affiliation{Minnesota Institute for Astrophysics, University of Minnesota, Minneapolis, MN 55455, USA}
\author{K. Lau}
\affiliation{Department of Physics, California Institute of Technology, Pasadena, CA 91125, USA}
\author{M. Lautzenhiser}
\affiliation{Department of Physics, University of Cincinnati, Cincinnati, OH 45221, USA}
\author{T. Liu}
\affiliation{Department of Physics, Stanford University, Stanford, CA 94305, USA}
\author{S. C. Mackey}
\affiliation{Kavli Institute for Cosmological Physics, University of Chicago, Chicago, IL 60637, USA}
\affiliation{Department of Physics, Astronomy \& Astrophysics, Enrico Fermi Institute, University of Chicago, Chicago, IL 60637, USA}
\author{N. Maher}
\affiliation{School of Physics and Astronomy, University of Minnesota, Minneapolis, MN 55455, USA}
\author{K. G. Megerian}
\affiliation{Jet Propulsion Laboratory, California Institute of Technology, Pasadena, CA 91109, USA}
\author{L. Minutolo}
\affiliation{Department of Physics, California Institute of Technology, Pasadena, CA 91125, USA}
\author{L. Moncelsi}
\affiliation{Department of Physics, California Institute of Technology, Pasadena, CA 91125, USA}
\author{Y. Nakato}
\affiliation{Department of Physics, Stanford University, Stanford, CA 94305, USA}
\author{H. T. Nguyen}
\affiliation{Department of Physics, California Institute of Technology, Pasadena, CA 91125, USA}
\affiliation{Jet Propulsion Laboratory, California Institute of Technology, Pasadena, CA 91109, USA}
\author{R. O'Brien}
\affiliation{Department of Physics, California Institute of Technology, Pasadena, CA 91125, USA}
\affiliation{Jet Propulsion Laboratory, California Institute of Technology, Pasadena, CA 91109, USA}
\author{S. N. Paine}
\affiliation{Center for Astrophysics, Harvard \& Smithsonian, Cambridge, MA 02138, USA}
\author{A. Papen}
\affiliation{Department of Physics, Astronomy \& Astrophysics, Enrico Fermi Institute, University of Chicago, Chicago, IL 60637, USA}
\author{A. Patel}
\affiliation{Department of Physics, California Institute of Technology, Pasadena, CA 91125, USA}
\author{M. A. Petroff}
\affiliation{Center for Astrophysics, Harvard \& Smithsonian, Cambridge, MA 02138, USA}
\author{A. R. Polish}
\affiliation{Center for Astrophysics, Harvard \& Smithsonian, Cambridge, MA 02138, USA}
\affiliation{Department of Physics, Harvard University, Cambridge, MA 02138, USA}
\author{N. Precup}
\affiliation{School of Physics and Astronomy, University of Minnesota, Minneapolis, MN 55455, USA}
\author{T. Prouve}
\affiliation{Service des Basses Temperatures, Commissariat a l'Energie Atomique, 38054 Grenoble, France}
\author{C. Pryke}
\affiliation{School of Physics and Astronomy, University of Minnesota, Minneapolis, MN 55455, USA}
\author{C. D. Reintsema}
\affiliation{National Institute of Standards and Technology, Boulder, CO 80305, USA}
\author{T. Romand}
\affiliation{Department of Physics, California Institute of Technology, Pasadena, CA 91125, USA}
\author{M. Salatino}
\affiliation{Department of Physics, Stanford University, Stanford, CA 94305, USA}
\author{A. Schillaci}
\affiliation{Department of Physics, California Institute of Technology, Pasadena, CA 91125, USA}
\author{B. Schmitt}
\affiliation{Center for Astrophysics, Harvard \& Smithsonian, Cambridge, MA 02138, USA}
\author{C. D. Sheehy}
\affiliation{School of Physics and Astronomy, University of Minnesota, Minneapolis, MN 55455, USA}
\author{B. Singari}
\affiliation{School of Physics and Astronomy, University of Minnesota, Minneapolis, MN 55455, USA}
\affiliation{Minnesota Institute for Astrophysics, University of Minnesota, Minneapolis, MN 55455, USA}
\author{A. Soliman}
\affiliation{Department of Physics, California Institute of Technology, Pasadena, CA 91125, USA}
\affiliation{Jet Propulsion Laboratory, California Institute of Technology, Pasadena, CA 91109, USA}
\author{T. St. Germaine}
\affiliation{Center for Astrophysics, Harvard \& Smithsonian, Cambridge, MA 02138, USA}
\author{A. Steiger}
\affiliation{Department of Physics, California Institute of Technology, Pasadena, CA 91125, USA}
\author{B. Steinbach}
\affiliation{Department of Physics, California Institute of Technology, Pasadena, CA 91125, USA}
\author{R. Sudiwala}
\affiliation{School of Physics and Astronomy, Cardiff University, Cardiff, CF24 3AA, United Kingdom}
\author{G. Teply}
\affiliation{Department of Physics, California Institute of Technology, Pasadena, CA 91125, USA}
\author{K. L. Thompson}
\affiliation{Department of Physics, Stanford University, Stanford, CA 94305, USA}
\affiliation{Kavli Institute for Particle Astrophysics and Cosmology, Stanford University, Stanford, CA 94305, USA}
\author{C. Tucker}
\affiliation{School of Physics and Astronomy, Cardiff University, Cardiff, CF24 3AA, United Kingdom}
\author{A. D. Turner}
\affiliation{Jet Propulsion Laboratory, California Institute of Technology, Pasadena, CA 91109, USA}
\author{C. Verg\`{e}s}
\affiliation{Physics Division, Lawrence Berkeley National Laboratory, Berkeley, CA 94720, USA}
\author{A. G. Vieregg}
\affiliation{Kavli Institute for Cosmological Physics, University of Chicago, Chicago, IL 60637, USA}
\affiliation{Department of Physics, Astronomy \& Astrophysics, Enrico Fermi Institute, University of Chicago, Chicago, IL 60637, USA}
\author{A. Wandui}
\affiliation{Department of Physics, California Institute of Technology, Pasadena, CA 91125, USA}
\author{A. C. Weber}
\affiliation{Jet Propulsion Laboratory, California Institute of Technology, Pasadena, CA 91109, USA}
\author{J. Willmert}
\affiliation{School of Physics and Astronomy, University of Minnesota, Minneapolis, MN 55455, USA}
\author{C. L. Wong}
\affiliation{Center for Astrophysics, Harvard \& Smithsonian, Cambridge, MA 02138, USA}
\affiliation{Department of Physics, Harvard University, Cambridge, MA 02138, USA}
\author{W. L. K. Wu}
\affiliation{Department of Physics, California Institute of Technology, Pasadena, CA 91125, USA}
\affiliation{Kavli Institute for Particle Astrophysics and Cosmology, Stanford University, Stanford, CA 94305, USA}
\affiliation{SLAC National Accelerator Laboratory, Menlo Park, CA 94025, USA}
\author{H. Yang}
\affiliation{Department of Physics, Stanford University, Stanford, CA 94305, USA}
\author{C. Yu}
\affiliation{Kavli Institute for Cosmological Physics, University of Chicago, Chicago, IL 60637, USA}
\affiliation{Argonne National Laboratory, High Energy Physics Division, Lemont, IL 60439, USA}
\author{L. Zeng}
\affiliation{Center for Astrophysics, Harvard \& Smithsonian, Cambridge, MA 02138, USA}
\author{C. Zhang}
\affiliation{Kavli Institute for Particle Astrophysics and Cosmology, Stanford University, Stanford, CA 94305, USA}
\author{S. Zhang}
\affiliation{Department of Physics, California Institute of Technology, Pasadena, CA 91125, USA}

\begin{abstract}

The South Pole is among the driest sites on Earth, and the resulting high atmospheric transparency makes it an excellent site for millimeter-wave astronomical observations.  Nevertheless, for bolometric surveys of the cosmic microwave background (CMB), fluctuations in line-of-sight water vapor drive variable atmospheric emission, which has a significant impact on survey sensitivity. We have deployed a water vapor radiometer (WVR) to characterize these fluctuations in space and time and to compare atmospheric and CMB signals with other sites.  This WVR has been monitoring conditions from the South Pole over the past nine years, co-located and co-observing with the BICEP series of telescopes. In this work, we analyze the performance of the WVR and the characteristics of the South Pole atmosphere and demonstrate coherent wind-driven transport of water vapor structures across the site.  We show that fluctuations in CMB detector timestreams are well correlated with fluctuations in atmospheric water vapor, providing independent justification for baseline filtering procedures used to remove these fluctuations and providing a new pathway to detect and mitigate the temperature-to-polarization leakage that currently limits the precision of ground-based CMB polarization measurements.



\end{abstract}

\newcommand\sectionprelude{
  \vspace{1em}
}

\section{Introduction}
The geographic South Pole is widely recognized as one of the most favorable locations on Earth for millimeter and submillimeter observations due to its high elevation, low atmospheric water vapor, and stable atmosphere free from diurnal variations. In particular, the exceptional stability of the atmosphere \citep{Bussmann_2005} and continuous access to regions of the sky with low celestial foregrounds make it an ideal site for studying the Cosmic Microwave Background (CMB), critical to understanding the early history of the Universe by testing inflationary theories \citep{Guth_InflationaryUniverse}.
The BICEP telescopes, deployed at the South Pole, have progressively tightened constraints on inflation over the past decade \citep{BK18}.
For the BICEP telescopes (and any ground-based instrument), the observing bandpasses are carefully tuned to match atmospheric transmission windows, that is frequency ranges that lie between strong molecular lines of water vapor ($H_{2}O$) and oxygen ($O_2$) \citep{Hui_BicepArrayPolarimeter}. Observing in these windows minimizes atmospheric emission noise and its fluctuations, which are the dominant constraints on observing efficiency.\newline
Water vapor is the most important, and the most variable, atmospheric constituent affecting CMB observations.
A water vapor radiometer (WVR) capable of scanning in azimuth and elevation was originally deployed at the South Pole to characterize atmospheric conditions and to facilitate direct comparisons with other high-altitude dry sites, such as the high Atacama Desert in Chile \citep{Morris2025,mackey2026}, Greenland ice sheet, or the Tibetan plateau, which are also considered for precision cosmological measurements \citep{barkats2018}. In this work, we demonstrate that the WVR data effectively reveal the spatiotemporal characteristics of the South Pole atmosphere.
We also show that atmospheric data acquired with the WVR correlate well with contemporaneous CMB data acquired by the Keck telescopes, which are part of the BICEP series of CMB polarization telescopes. 
This indicates that the scanning WVR data are an effective tool for comparing the quality of CMB observing sites and confirms that atmospheric background subtraction procedures used in CMB data processing have a solid physical grounding.\newline
This paper is organized as follows. Section 2 describes the WVR instrument, observing strategy, data acquisition, and instrumental systematics, including polarization-dependent mirror loss. In Section 3, we present the analysis of the precipitable water vapor (PWV) distribution and its temporal and spatial fluctuations. Section 4 characterizes the correlation between atmospheric fluctuations measured by the WVR and those observed in CMB data from the Keck receivers. Section 5 summarizes our conclusions and discusses prospects for future improvements in atmospheric modeling and systematics characterization using data acquired with a radiometer operating independently from the main CMB receiver.


\section{The Water Vapor Radiometer}
\subsection{Instrument Design and Scanning Strategy}

The WVR consists of a multichannel radiometer, originally developed for the ALMA observatory by Omnisys Instruments \citep{alma1-Emrich2009,alma2-Nikolic2013}, and a two-mirror scanning periscope which couples this instrument to the sky \citep{barkats2018}. The radiometer is an uncooled heterodyne receiver that observes the sky in 4 double-sideband (DSB) spectral channels centered on the 183.31 GHz water vapor emission line.  These channels are defined by intermediate frequency (IF) filters with bandwidths ranging from 1.5 -- 2.5 GHz and with center frequencies arranged to achieve spectral coverage from 175 -- 192 GHz. Channel 0 is closest to the center of the water vapor line, while channel 3 is the furthest.

\begin{figure}[htb!]
    \centering
    \includegraphics[width=0.85\columnwidth]{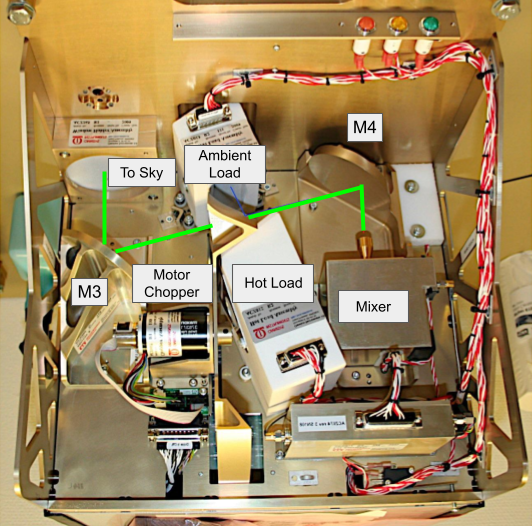}
    \caption{Picture of the RF side of the radiometer system showing hot load, ambient load, and mirrors. Each full rotation of the chopper mirror mount corresponds to a measurement sequence of [hot load, sky view A, ambient load, sky view B].}
    \label{fig:wvr_baseunit_schematics}
\end{figure}
\begin{figure}[htb!]
    \centering
    \includegraphics[width=1.\columnwidth]{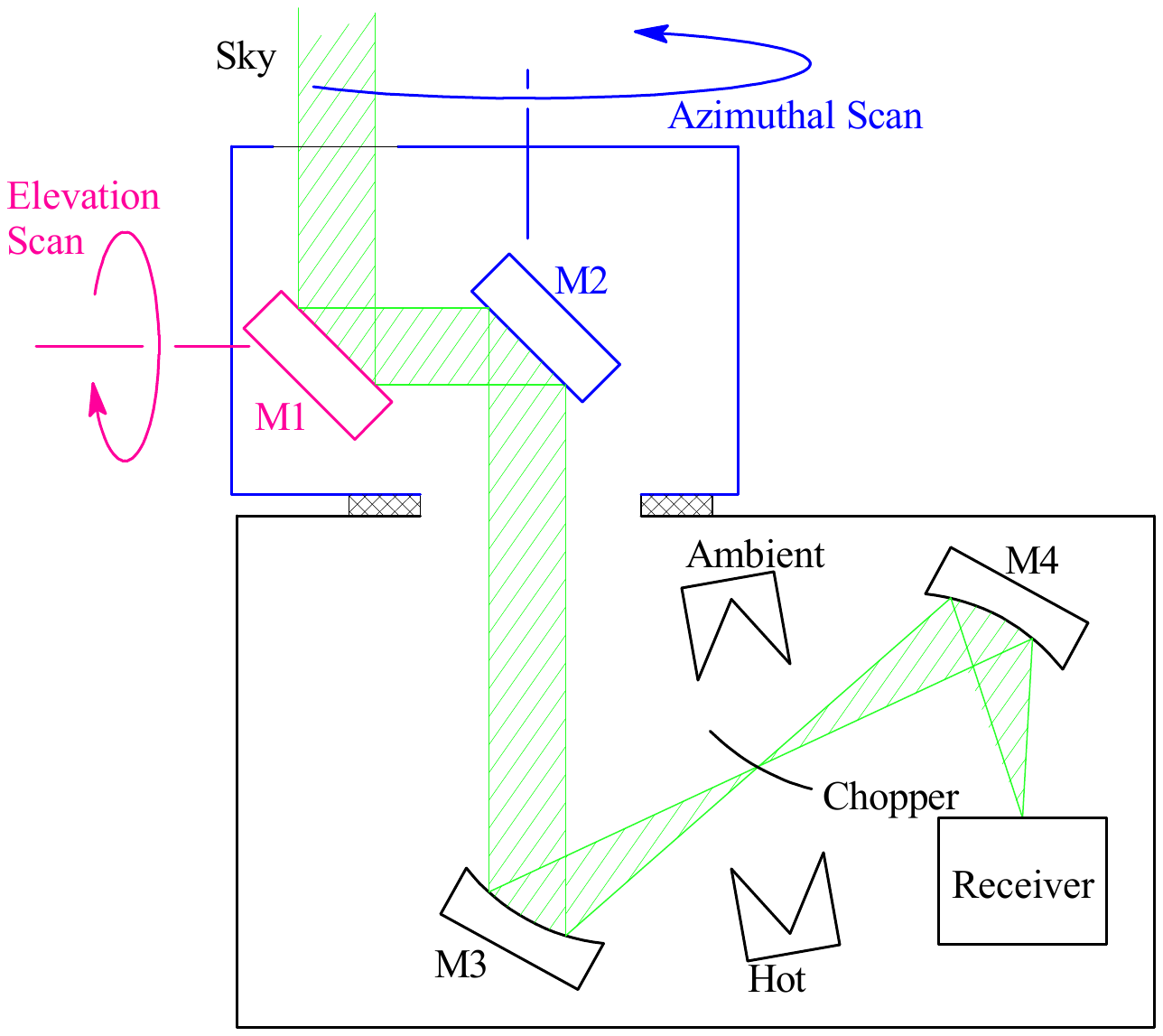}
    \caption{Mirror configuration in the water vapor radiometer. M1 and M2 are the flat mirrors in the rotation stage that rotate as a unit during an Az scan. M3 and M4 are curved mirrors within the receiver front end. Hot and Ambient indicate calibration load mirrors on the rotating chopper mirror mount. The black box outline corresponds to the receiver unit shown in Fig.\ref{fig:wvr_baseunit_schematics}. Under normal operation, the blue outlined portion scans in azimuth, while the magenta mirror scans in elevation, with azimuthal rotation disabled, to perform sky dips.}
    \label{fig:wvr_mirror_scheme}
\end{figure}

The heterodyne receiver (Fig. ~\ref{fig:wvr_baseunit_schematics}) was designed to be operated in the temperature regulated environment of an ALMA antenna receiver cabin, between $16 \degree$C and $22 \degree$C. To operate it at the South Pole (temperatures between $-20 \degree$C and $-75 \degree$C), it was housed in a custom enclosure, whose air temperature is maintained using an analog PID temperature-controlled heater stage and an impeller fan. 
In addition to maintaining a positive pressure inside the environmental enclosure, the impeller fan draws in a small amount of outside air through variable size openings.
The radiometer input port views the sky through a periscope (Fig.~\ref{fig:wvr_mirror_scheme}) consisting of two flat aluminum mirrors that enable scanning the sky view continuously over 360$\degree$ in azimuth and from 10$\degree$ to 90$\degree$ in elevation. 
 Internally, the radiometer optics include two curved mirrors on a rotating mount that can direct the view toward respective hot (363.15 K) and ambient (283.15 K) calibration loads.  The load temperatures are regulated to better than 10 mK stability \citep{alma2-Nikolic2013}.
The chopping mirror rotates at approximately 5 Hz, sequentially viewing the sky, the ambient load, the sky again, and the hot load. 

Each integration lasts about 0.048 s per position, and all four frequency channels are read simultaneously.
\begin{figure}[htb!]
	\centering
	\includegraphics[width=1\columnwidth]{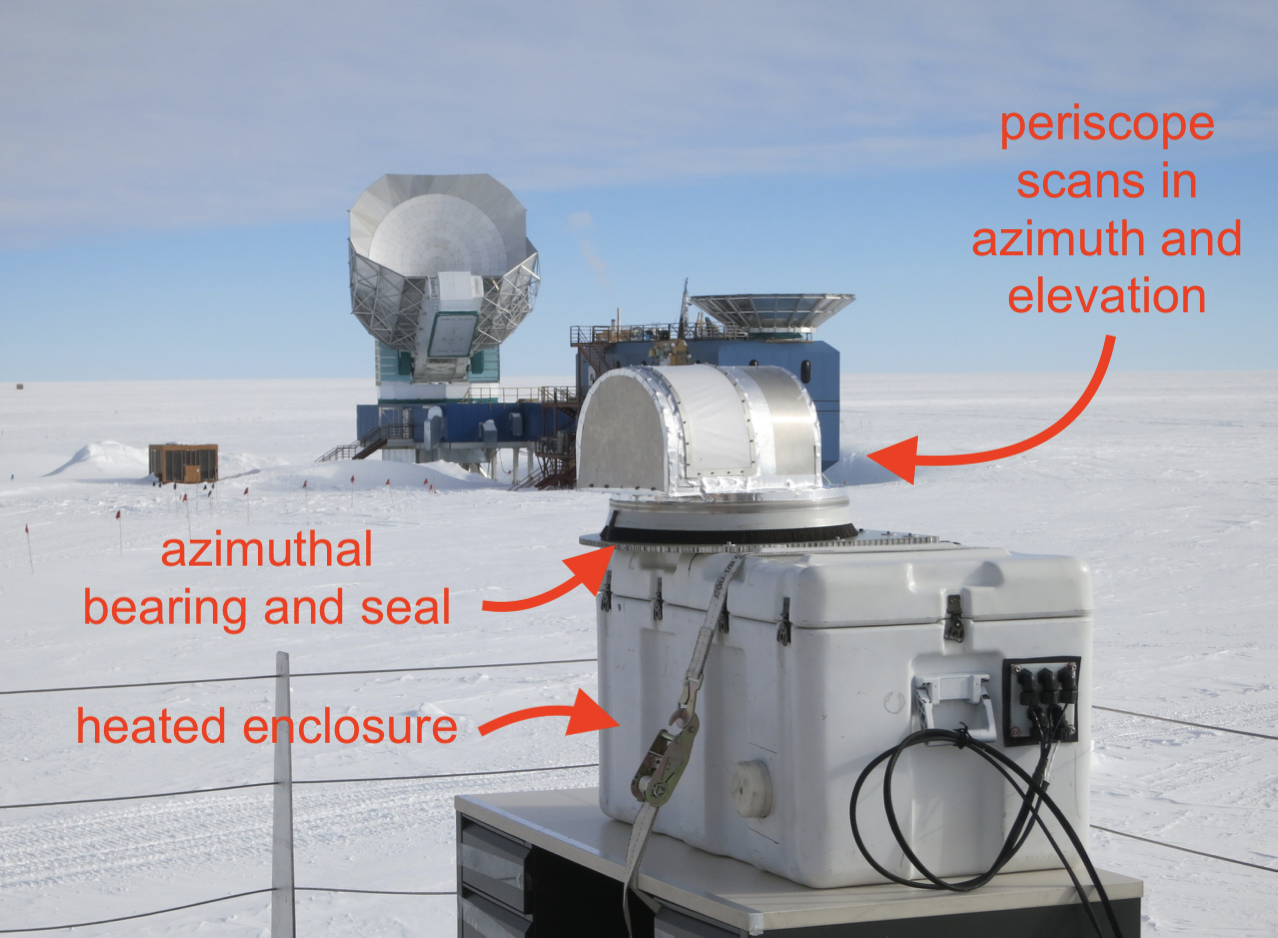}
	\caption{A picture of the WVR enclosure located on the roof of MAPO,
the same building where BICEP Array is located at the South Pole Station.
The main box is the enclosure for the system shown in Fig.~\ref{fig:wvr_baseunit_schematics}.
The ‘mailbox’ with the curved environmental window houses the periscope mirrors that enable the WVR to scan in azimuth and elevation.}
	\label{fig:wvr_picture}
\end{figure}

The WVR scanning strategy consists of observing the sky continuously in azimuth at a speed of 12$\degree$/s and at an elevation of 55$\degree$, chosen to match the elevation at the center of the BICEP CMB maps \citep{BK18}.
Additionally, the radiometer performs a sky dip, which is an elevation scan during which the radiometer moves from 55$\degree$ up to 90$\degree$, then down to 15$\degree$, then back to 55$\degree$, once per hour to acquire a direct measurement of the atmospheric brightness temperature versus elevation.  Except for maintenance, upgrades, and beam mapping operations, the WVR has been scanning the sky continuously since 2016 (Fig.~\ref{fig:wvr_picture}).


\subsection{Data processing and calibration}
\label{subsec: data_format}

We extract radiometric temperatures from data files containing a set of successive measurements pointing at the sky and at two loads. 
Data consist of successive measurements corresponding to hot load, sky position A, ambient load, sky position B, sampled at a cadence of 0.056 s.
Raw voltage readings are calibrated into brightness temperature units at the plane of the chopper using two reference temperatures:
\vspace{-.4em}
\begin{equation}
 \begin{split}   
  G=\frac{V_{\rm Hot}-V_{\rm Amb}}{T_{\rm Hot}-T_{\rm Amb}} 
    \end{split}
\end{equation}
\begin{equation}
 \begin{split}   
  T_{\rm Sky}=T_{\rm Amb}+\frac{V_{\rm Sky}-V_{\rm Amb}}{G}
\end{split}
\end{equation}
\vspace{-.5em}
\newline
For all the analysis that follows, the $V_{sky}$ at a given time/position will be the average of the two A/B measurements. The calibration gain 
$G$ is derived for each full 55 minute dataset, rather than estimated independently for each individual reading. This approach reduces the variance in the calibrated temperature measurements by a factor of two.
Throughout this paper, atmospheric brightness fluctuations are reported in Rayleigh---Jeans temperature units ($K_{RJ}$), derived by calibrating detector voltages against a load of known temperature, as explained above, and applying the Rayleigh---Jeans approximation to the Planck function. This convention, standard in CMB instrumentation, facilitates comparison across instruments and frequency bands.

\subsection{Beam Shape and Pointing}
\label{subsec:beam_and_pointing}

The WVR beam and pointing parameters are derived from beam maps obtained using the Sun as a point source during the austral summer. 

\begin{figure}[htb!] 
  \centering
\includegraphics[width=.8\columnwidth]{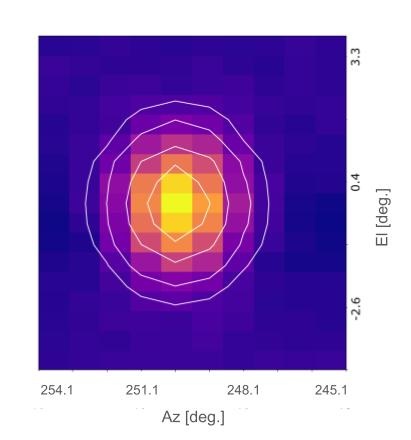}
  \caption{Peak Normalized median of the beam maps in the four radiometer channels after background subtraction. Contours from the 2D Gaussian best-fit model are shown. 
  }
  \label{fig:beam_map_2DGauss_contours}
  \end{figure}

\begin{figure*}[t!]
  \centering
  \includegraphics[width=\textwidth]{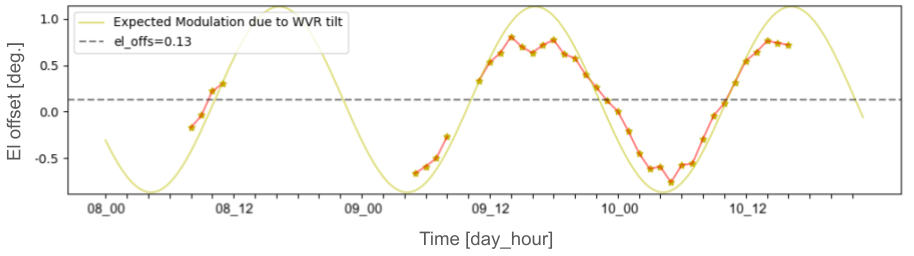}
  \caption{
  Elevation offset extracted from the 2018 beam-mapping campaign. The full dataset was acquired over three days in December 2018; the x-axis shows time in the format $day\_hour$. The modulation in $El_{off}$ is generated by a physical tilt of the radiometer. In Sun-centered coordinates, this effect appears as a 24-hour azimuthal modulation due to the changing solar position in azimuth. Missing data points correspond to datasets for which the fit failed.
  }
  \label{fig:beam_parameters_2018}
\end{figure*}

Each observation campaign includes the acquisition of approximately 40 beam maps. Each beam map consists of 17 full 360$^\circ$ azimuth scans taken at various elevations, ensuring comprehensive coverage of the Sun. The elevation step between scans is $0.25 \degree$. From each beam map, estimates of FWHM$_{Az}$ and FWHM$_{El}$, and pointing offsets are extracted. The final seasonal values are determined by averaging these estimates over the entire set of maps.
The extraction of beam and pointing parameters begins by independently generating Sun maps for each of the four radiometer channels. These maps are then transformed from WVR coordinates to Sun-centered coordinates using the relations $Az_{off} =  Az_{wvr} - Az_{sun}$ and $El_{off} = El_{wvr}-El_{sun}$. 
The background, which includes various systematic effects detailed in Sec.\ref{subsec:wvr_systematics}, is subtracted from each Sun-centered map. Next, the median is computed on the four cleaned channel maps. 
Finally, a 2D Gaussian fit is applied to determine $Az_{off}$, $El_{off}$, FWHM$_{Az}$, and FWHM$_{El}$. Here, the offset parameters correspond to the coordinates of the Gaussian center, while the beam size is related to the Gaussian width via $\sigma$ by FWHM $ = 2 \sigma \sqrt{ln(2)}$ (Fig.~\ref{fig:beam_map_2DGauss_contours}).
The final (seasonal) value for FWHM$_{Az}$ and FWHM$_{El}$ is picked by averaging over the full set of maps (Fig.~\ref{fig:beam_parameters_2018}). 
From the 2018 beam mapping campaign, we measured a circular beam (with an ellipticity of $\sim 1\%$), of size FWHM$_{Az} = (2.73 \pm 0.03) \degree $ and FWHM$_{El} = (2.85 \pm 0.1)\degree$. 
The pointing parameters are $Az_{off}=(250 \pm 0.2)\degree$ and $El_{off}=(0.13 \pm 0.1)\degree$. Furthermore, $El_{off}$ is modulated in Az with a time period of $24$ hrs; the cause is to be attributed to a physical tilt in the radiometer platform, and its effects on the data will be described in Sec. \ref{subsec:wvr_systematics}.

\subsection{Instrumental systematics}
\label{subsec:wvr_systematics}

The WVR data present systematic effects that require correction. 
In this section, we describe the nature of these effects and outline the methodology used to model and remove them from the data.
To visualize the temporal and spatial structure of the atmosphere, we construct atmograms. Each atmogram is a stack of consecutive azimuth scans, with azimuth on the vertical axis and time on the horizontal axis. Each pixel represents the measured sky brightness temperature (or PWV) at that azimuth and time. This format allows us to track atmospheric fluctuations and identify structures moving across the field of view.

\begin{figure}[htb!]
    \centering
    \includegraphics[width=1.\columnwidth]{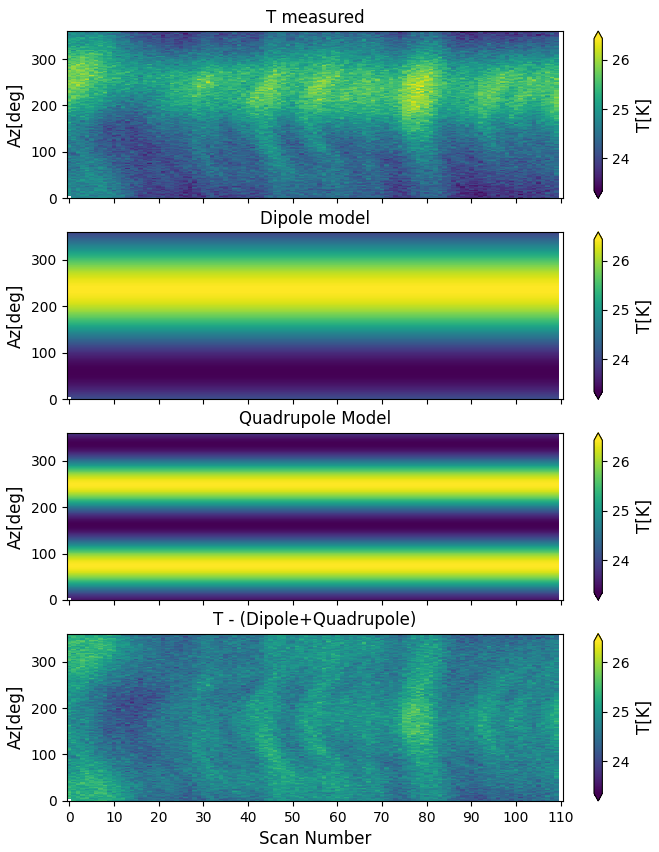}
	\caption{Sky temperature as measured by channel 0 (where the dipole signal is stronger) reported in the atmogram format, that is a horizontal time series of $360 \degree$ Az scans, for a total of 55 minutes of data. (a) $T_{sky}$ measured by channel 0 for a full Az scan. (b) Measured instrumental Dipole. (c) Measured instrumental Quadrupole. (d) $T_{sky}$ measured by channel 0 after removing the instrumental dipole and quadrupole signal.}
	\label{fig:quadrupole_T0quad}
\end{figure}

\subsubsection{Ground-Fixed Dipole in Azimuth}
\label{subsubsec: dipole}
\begin{figure*}[htb!]
\begin{center}
    \includegraphics[width=.9\textwidth]{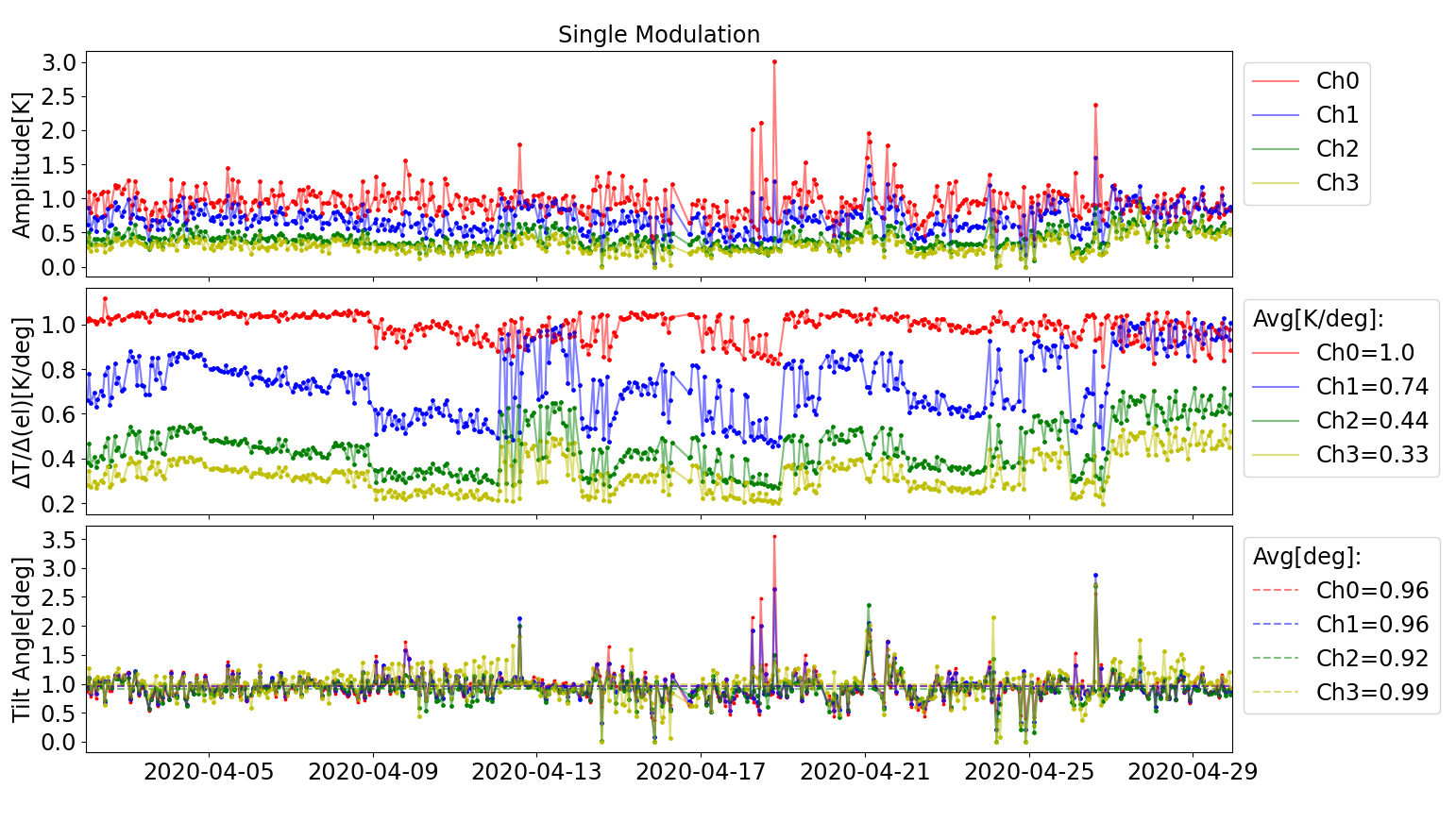}
  \caption{
  Dipole Amplitude, $dT/dEl$, and tilt angle for one month-long dataset. Top: Amplitude of the ground-fixed dipole signal for each of the four frequency channels. Middle: $dT/dEl$ for the same dataset extracted from sky dips. Bottom: Tilt angle for the same dataset extracted from the dipole amplitude using the conversion factors shown in the middle panel.
  }
  \label{fig:tilt_angle}
  \end{center}
\end{figure*}

The elevation offset extracted from beam map observations shows a modulation over a time scale of $\sim1$ day, which is consistent with a physical tilt from horizontal in the WVR. This tilt generates a ground-fixed dipole signal in azimuth. 
The amplitude of this signal fluctuates in time, and it is larger in channel 0 and smaller in channel 3 (Fig.~\ref{fig:tilt_angle}), as expected for a signal generated by a tilt from the horizontal that causes a change of atmospheric emission load while scanning in azimuth.
In the data cleaning process, the signal removed from each azimuth scan is a sinusoid, where the amplitude corresponds to the monthly averaged tilt---converted back into temperature using the appropriate $\frac{dT}{dEl}$ at the azimuth scan elevation angle ($55\degree$) (Fig.~\ref{fig:tilt_angle}). The phase of the sinusoid is the average over the full 110 scans in one atmogram.
The resulting tilt angle from the 2020 dataset is approximately $\sim 1 \degree$, which is consistent with the elevation offset modulation observed in beam maps acquired in 2018 (Fig.~\ref{fig:beam_parameters_2018}).
We expect the tilt amplitude and direction to remain stable over timescales of several months, as they are primarily driven by the slow mechanical tilting of the building. This drift occurs at a rate of a few millidegrees per year, consistent with long-term pointing measurements from BICEP and Keck.

\vspace{-.6em}

\subsubsection{Ground-Fixed Quadrupole in Azimuth}
\label{subsubsec: quadrupole}

Data also show a ground-fixed quadrupole term in azimuth.
The amplitude of the quadrupole is $\sim 0.5 K_{RJ}$, and it is extremely stable in time. Different from the dipole term, this signal is stronger in channel 3 and weaker in channel 0, suggesting it is not generated by the change in atmospheric load while the radiometer is scanning in azimuth.
This quadrupole effect is consistent with the effect of polarized reflectivity of the $45 \degree$ flat mirrors in the WVR optics. 
Table \ref{tab:quadrupole_amplitude} presents measured quadrupole amplitude for the 4 channels, averaged for the month of April  2020, as well as a predicted quadrupole amplitude based on a simple model, presented below.
The WVR has two curved mirrors and two flat mirrors at $45\degree$ to the optic axes (see Fig.~\ref{fig:wvr_mirror_scheme}). The two flat mirrors, M1 and M2, rotate in azimuth as a unit to perform a $360\degree$ azimuthal scan. During each scan the polarization direction makes a full $360\degree$ rotation on the surface of each of the two mirrors. On the curved mirrors, M3 and M4, the polarization direction does not change. 
All mirrors are fabricated in aluminum, with an electrical resistivity $\sigma = 2.65\ \mu \Omega$ cm \citep{Al_resistivity} and a skin depth $\delta = 0.19$ $\mu$m at $f = 183$ GHz. Therefore the mirror effective surface resistance is:
\begin{equation}
    R_s = \frac{1}{\sigma \delta} = 0.125 \ \Omega / \square
    \label{Eq: Rs}
\end{equation}
At normal incidence the E and H field of the radiation are related by the relation $E_0/H_0 = 120 \pi \Omega / \square$. Upon reflection E cancels, and H at the mirror surface is $H_{\parallel}=2H_0$. This H generates a surface current that is perpendicular to $H_{\parallel}$ and gives rise to a voltage drop. 
Because the resistivity is non-zero (Eq. \ref{Eq: Rs}), this generates an electric field given by:
\vspace{-.3em}
\begin{equation}
    E_s = R_s H_{\parallel} = 2 R_s H_0
    \label{eq:E_s}
\end{equation}
\vspace{-0.3em}
In each of the two mirrors radiation losses are calculated from the Poynting vector:
\vspace{-0.3em}
\begin{equation}
    \epsilon_M = \frac{E_s \times R_{\parallel}}{E_0 \times H_0} = \frac{4 R_s}{120 \pi}
\end{equation}
\vspace{-0.3em}
This is also the mirror thermal emission coefficient, by reciprocity.
Therefore, the mirror emissivity changes the WVR zenith temperature by:
\vspace{-0.3em}
\begin{equation}
    \Delta T_M = \epsilon_M (T_M - T_{ATM})
    \label{eq: dTm}
\end{equation}
\vspace{-0.3em}
Taking into account that there are two mirrors, the total variation on the zenith temperature will be $2 \Delta T_M$.

\begin{figure}[htb!]
    \centering
    \includegraphics[width=0.9\columnwidth]{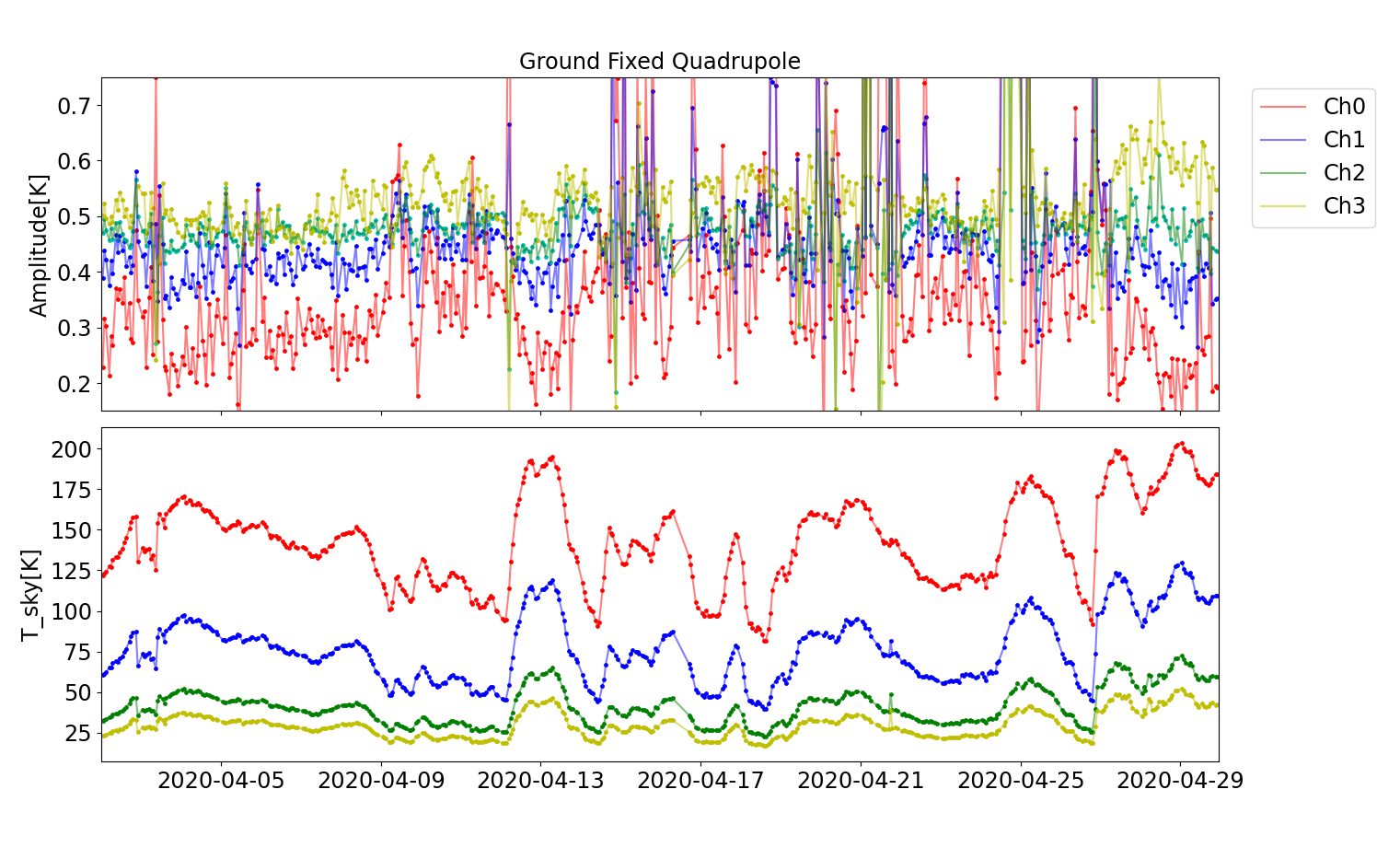} 
    \newline
    \caption{Amplitude of the ground-fixed quadrupole signal for each of the four frequency channels and for a one month dataset. The amplitude of this signal is very stable with respect to temperature fluctuations, and it is higher in channel 3 and lower in channel 0.}
    \label{fig:quad_amp}
\end{figure}

Assuming a mirror temperature $T_M \sim 240$ K, Eq. \ref{eq: dTm} and $T_{atm}$ in table \ref{tab:quadrupole_amplitude} can be used to estimate the expected variation in the measured temperature. An estimate of the expected variation in each of the four channels is reported in table \ref{tab:quadrupole_amplitude}.

 The expected $dT$ agrees very well with the measured $dT$.

\begin{table}[htb!]
\begin{center}
\begin{tabular}{c c c c c}
  & Ch0 & Ch1 & Ch2 & Ch3 \\ 
 \hline
 $T_{atm}[K]$ & 150 & 75 & 40 & 25\\
 $\Delta T_{Q,expected} [K]$ & 0.24 & 0.44 & 0.53 & 0.57\\  
  $\Delta T_{Q,measured} [K]$ & 0.28 & 0.40 & 0.50 & 0.55\\  
 \end{tabular}
 \caption{Expected and measured amplitudes of the ground-fixed quadrupole signal for the four WVR channels, averaged over April 2020. $T_{\rm ATM}$ is the atmospheric brightness temperature, while $\Delta T_{Q,\rm expected}$ and $\Delta T_{Q,\rm measured}$ are the predicted and measured quadrupole amplitudes, respectively.}
  \label{tab:quadrupole_amplitude}
  \end{center}
 \end{table}

As the periscope mirrors turn, the parallel component of H varies by a factor of $\sqrt{2}$, and the losses scale as $H_{\parallel}^2$, so the losses are modulated by a factor of 2.
To clean azimuth scans from this signal we fit a $2 \theta$ sinusoid to each azimuth turn, constraining amplitude and phase to be constant over one full Az map (i.e. 110 azimuth scans).
For each one month dataset we extract the amplitude and the phase of this signal scan by scan, and we find the monthly average for each radiometer channel (Fig.~\ref{fig:quad_amp}).
The signal we finally remove from each azimuth turn is a $2 \theta$ 
sinusoid for which the amplitude and phase are the monthly averaged parameters.  The amplitude is channel dependent, while the phase in azimuth is not (Fig.~\ref{fig:quadrupole_T0quad}).

\section{The South Pole Atmosphere}
\label{sec:sp_atmosphere}

\subsection{From Radiometric Temperatures to Precipitable Water Vapor}
\label{subsec:T_to_PWV}

\begin{figure}[htb!] 
  \centering
    \includegraphics[width=1\columnwidth]{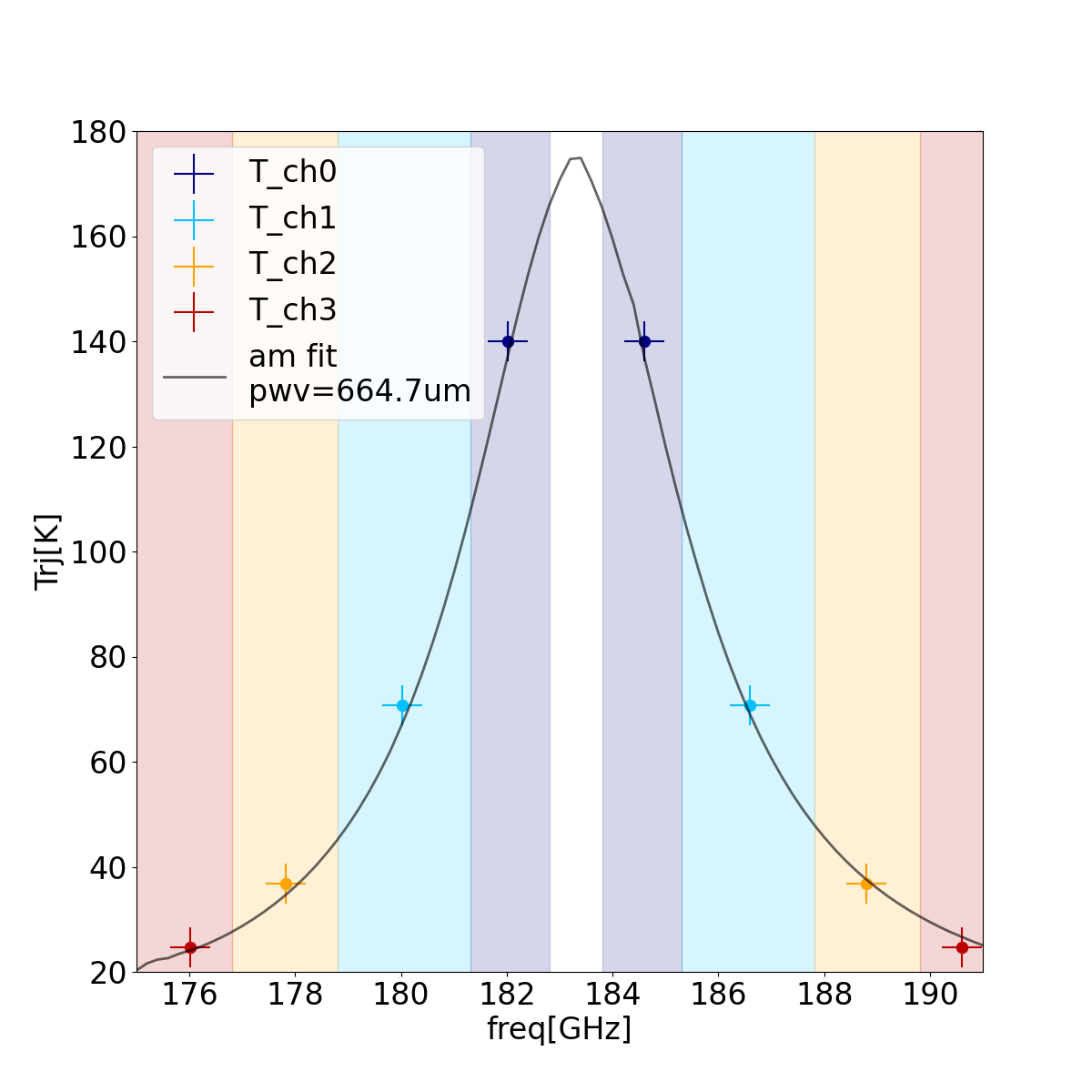}
  \caption{The plot shows a water line fit generated using am, applied to the four zenith temperatures from a sky dip measured by the radiometer's four double sideband channels. The shaded regions indicate the passbands of the four channels. The resulting PWV for this datapoint is $664.7 \mu$m. The datapoint was extracted from a sky dip acquired on 04/19/2020.} 
\label{fig: Am_line_shape}
\end{figure}

The four radiometer bandwidths and band centers were originally designed to optimally sample the shape of the 183 GHz water vapor emission line for typical conditions at the ALMA site (PWV $\lesssim $ 10 mm). 
Therefore, from each set of four radiometric temperatures, an estimate of precipitable water vapor can be extracted by fitting to the four temperatures the shape of the line (Fig. \ref{fig: Am_line_shape}). For this fit we use the $am$ atmospheric modeling software \citep{am_paine}.
Two important steps are required to ensure the accuracy of our final PWV estimate.
First, it is crucial to determine whether variations in radiometer temperatures are driven by changes in precipitable water vapor (PWV), atmospheric pressure ($P_{atm}$), atmospheric temperature ($T_{atm}$), or a combination of these three factors (see Sec. \ref{subsubsec:T_rad_vs_PWV_curves}).
Second, an appropriate model for the vertical structure of the atmosphere must be selected to accurately interpret the radiometric data (see Sec. \ref{subsubsec:atmo_models}).

\subsubsection{$P_{atm}$, $T_{atm}$ and PWV Mass Fraction Fluctuations}
\label{subsubsec:T_rad_vs_PWV_curves}

To determine whether variations in radiometric temperatures are caused by changes in atmospheric pressure ($P_{atm}$), atmospheric temperature ($T_{atm}$), precipitable water vapor PWV or a combination of these factors, we analyzed the relative zenith temperature variations in channels 1, 2, and 3 as a function of the temperature variation in channel 0, which is closest to the center of the water vapor line.

 \begin{figure}[htb!]
    \centering
    \includegraphics[width=1\columnwidth]{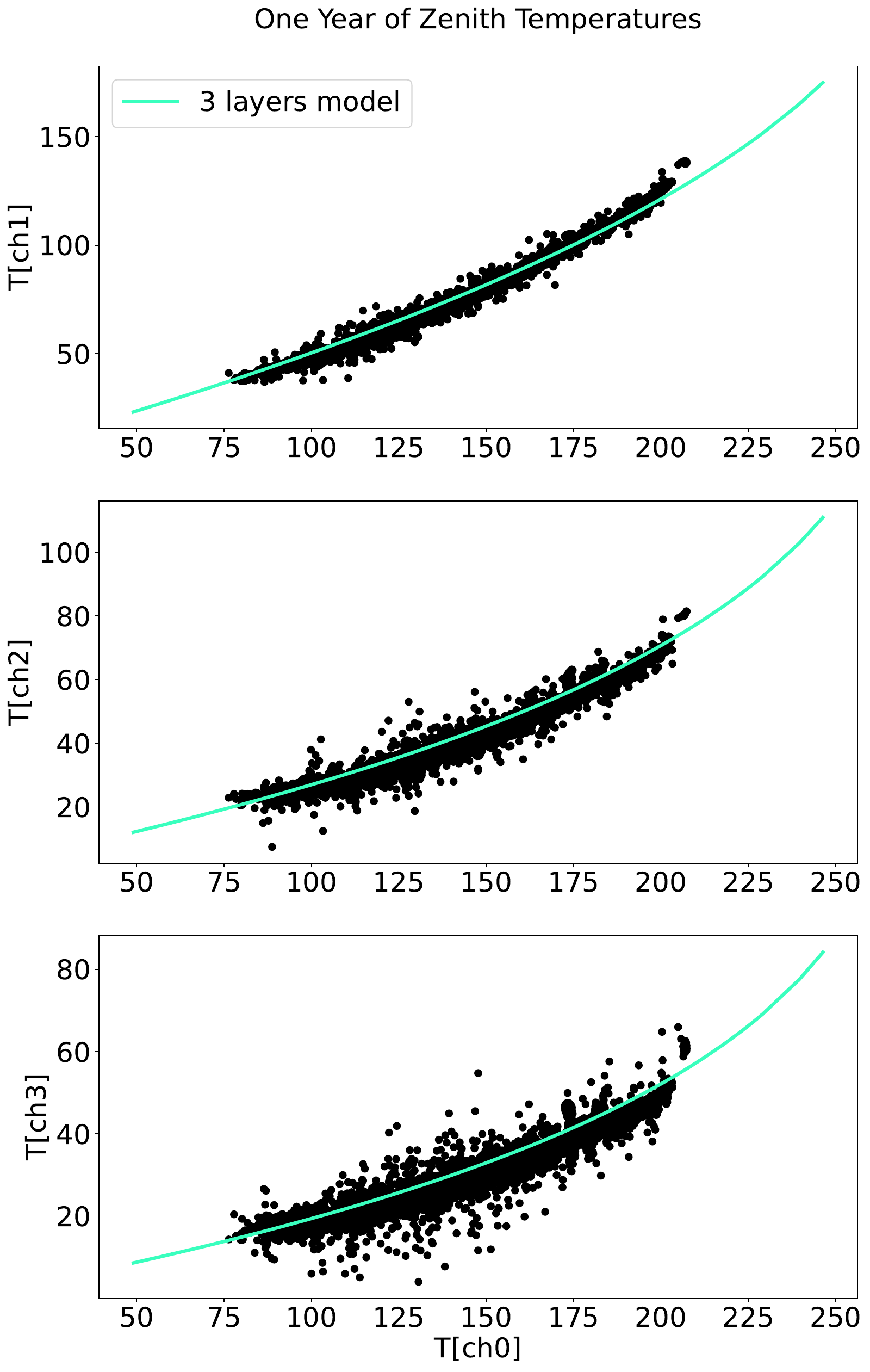} 
    \caption{Zenith temperatures recorded throughout 2020 by radiometer channels 1, 2, and 3 were plotted against the zenith temperatures measured by channel 0. 
    The data are well described by a three-layer atmospheric model in which the temperatures of the layers are held constant over different channels, while the optical depth $\tau$ is allowed to vary with time. Models with fewer than three thermal layers do not properly fit the data.
    The dataset includes one year of zenith temperature measurements derived from sky dip data.}

    \label{fig:scatterplots}
\end{figure}

We found that these curves are well described by a three-layer atmospheric model in which the layer temperatures ($T_{i}$) are held constant over different channels, while the layer opacities ($\tau_{i}$) are allowed to vary:

\begin{equation}
\begin{aligned}
 T_{ch}=T_{1}\cdot(1-{e^{-{\tau}_{1,ch}}}) \\
 + T_{2}\cdot e^{-\tau_{1,ch}} \cdot(1-e^{-{\tau}_{2,ch}}) \\
 + T_{3}\cdot e^{-(\tau_{1,ch}+\tau_{2,ch})} \cdot(1-e^{-{\tau}_{3,ch}})
\end{aligned}
\label{T_sky}
\end{equation}

where the transmission coefficients ${\tau_i}$ scale from channel to channel by a fixed proportionality factor.
\begin{figure*}[t!] 
  \centering
    \includegraphics[width=0.95\textwidth]{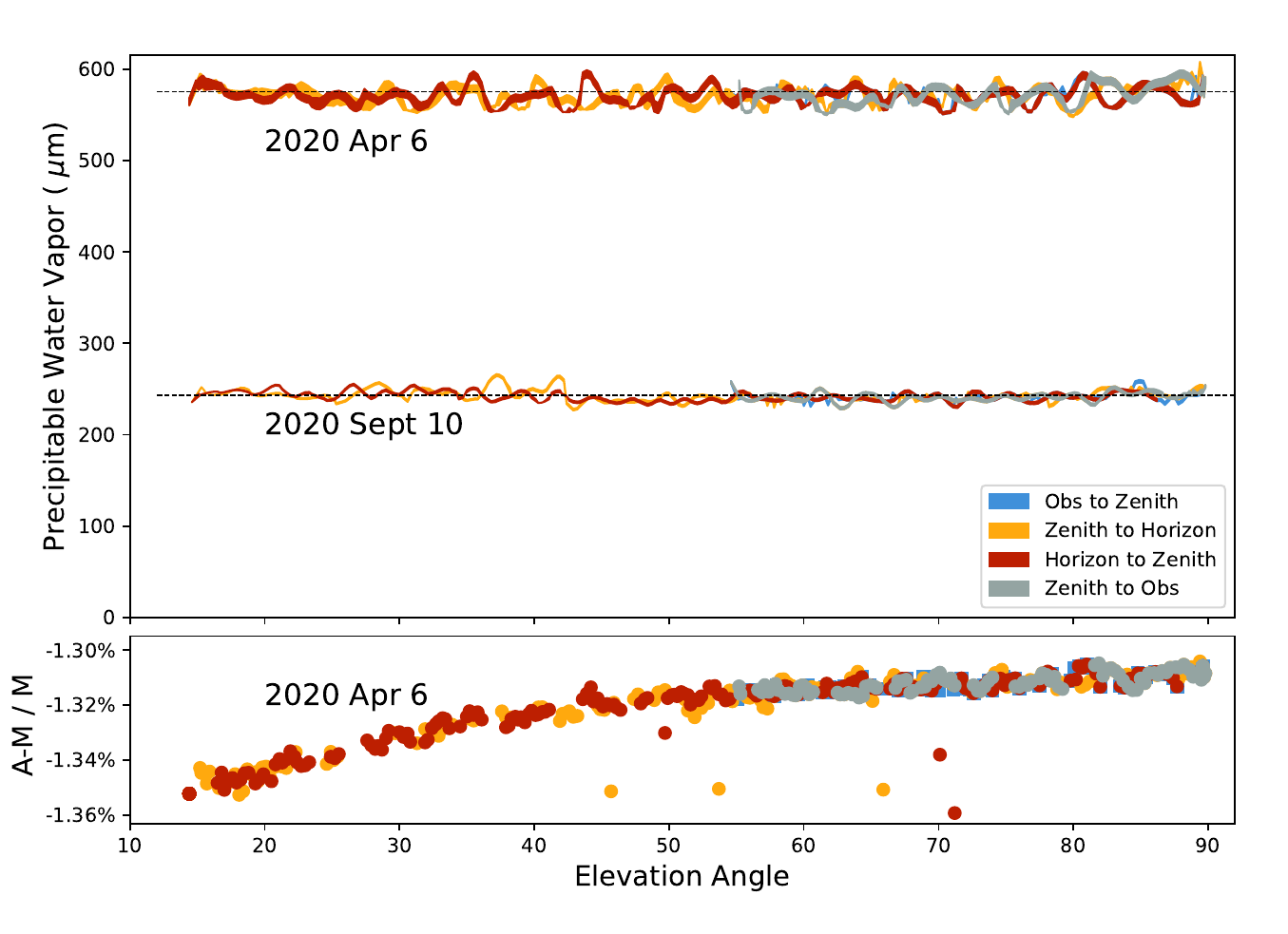}
  \caption{PWV as a function of elevation angle for two different sky dips acquired on separate days in 2020. Each dataset was fit using both the standard \texttt{am} South Pole annual average atmospheric template and a template constructed from contemporaneous MERRA data \citep{merra} specific to each date. The width of each trace encompasses the difference between the two fits. The bottom panel shows the percentage difference between the MERRA-based (M) and Annual model (A) fits as a function of elevation angle. The close agreement between the two approaches indicates that the Annual average atmospheric template provides sufficient accuracy for this analysis.}
\label{fig:PWV_variations_SkyDip}
\end{figure*}

This demonstrates that a simple three-layer atmospheric model with adjustable water vapor column densities and sensible priors on the layer temperatures can accurately model the relative channel brightness temperatures. Models with fewer than three thermal layers do not properly fit the data. Nevertheless, since detailed prior profiles are available from meteorological reanalysis data, we chose to use profile data from the MERRA-2 reanalysis to drive our PWV retrievals as discussed below.\newline

\subsubsection{A Multi-Layer Model for the Atmosphere}
\label{subsubsec:atmo_models}

A single-layer atmospheric model does not provide a good fit to the curves in Fig.~\ref{fig:scatterplots}, whereas a multi-layer atmospheric model does.
Am provides a 30-layer standard atmospheric model, in which each layer has a specific pressure and temperature and specific dry air, $H_{2}O$, and $O_3$ column densities. It also allows us to fit a single scale factor to the $H_{2}O$ column density in all the tropospheric layers. \newline
To evaluate whether a yearly averaged atmospheric template is sufficiently accurate, we constructed a second template based on daily pressure and temperature profiles derived from MERRA reanalysis data \cite{merra}.
Figure \ref{fig:PWV_variations_SkyDip} shows the zenith-projected PWV as a function of elevation for a representative sky dip observation. The plot compares PWV estimates obtained by fitting the WVR temperature data to both the yearly average am model and the date-specific MERRA-based template. The width of each trace in the plot encompasses the difference between the two fits.
The results demonstrate close agreement between the two approaches, indicating that the yearly average atmospheric template provides sufficient accuracy for our analysis.
Therefore, in the analysis that follows, all data are fitted using a 30-layer atmospheric model with seasonally averaged parameters ($P_{atm}, T_{atm}$).

\subsection{PWV Distribution}
\label{subsec:pwv_fluctuations_time}

\begin{figure}[htb!]
    \centering
    \includegraphics[width=1\columnwidth]{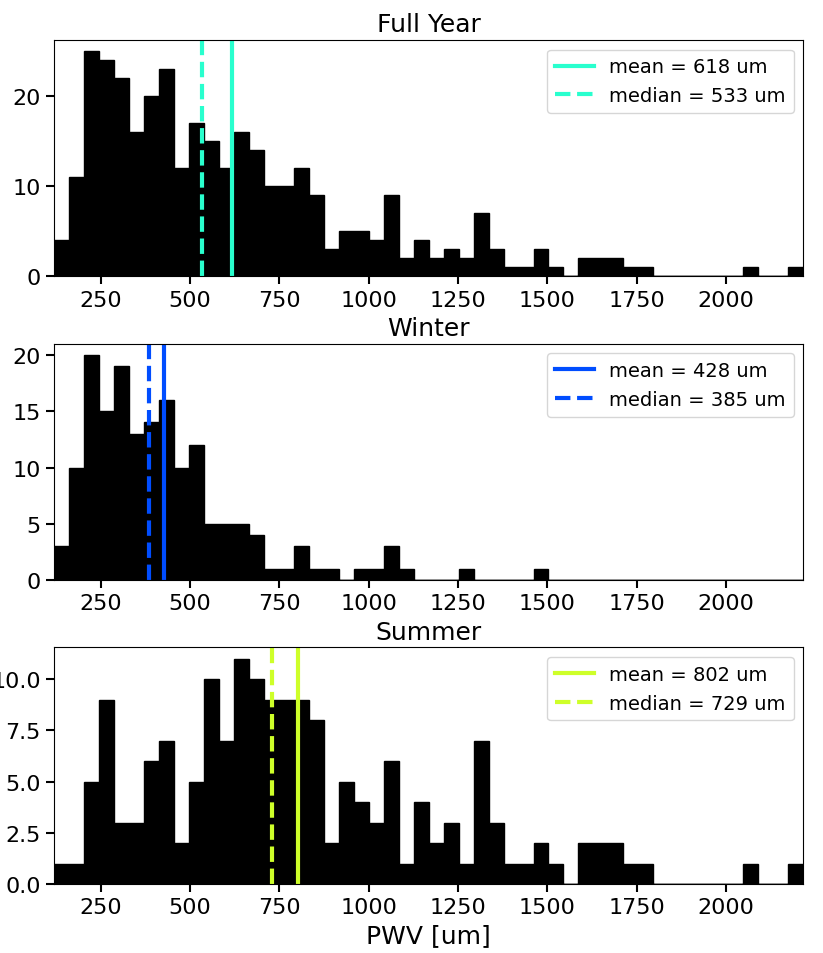} 
    \caption{Histograms reporting PWV data extracted from WVR zenith temperatures, for winter and summer separately and for the full year. Winter includes the months of Apr, May, Jun, Jul (to match the analysis done in \cite{Bussmann_2005}). Summer corresponds to the months of Nov, Dec, Jan, Feb. The mean and the median are reported for each of the three data samples.
    }
    \label{fig:sesonal_histo}
\end{figure}

We extracted one year of PWV data from sky dip measurements. The seasonal and annual PWV histograms (Fig.~\ref{fig:sesonal_histo}) indicate that the mean PWV for the austral winter of 2020 was 0.43 mm.
This value is less than half the mean PWV observed during the best six months at other internationally-important sites for millimeter observations, such as Mauna Kea (1.65 mm) and the Atacama Desert (1.00 mm) \citep{Bussmann_2005}. Combined with atmospheric stability and continuous access to the same sky at constant elevation, this extreme aridity makes the South Pole the premier site on Earth for deep cosmological surveys. \newline

\begin{figure}[htb!]
    \centering
    \includegraphics[width=1.\columnwidth]{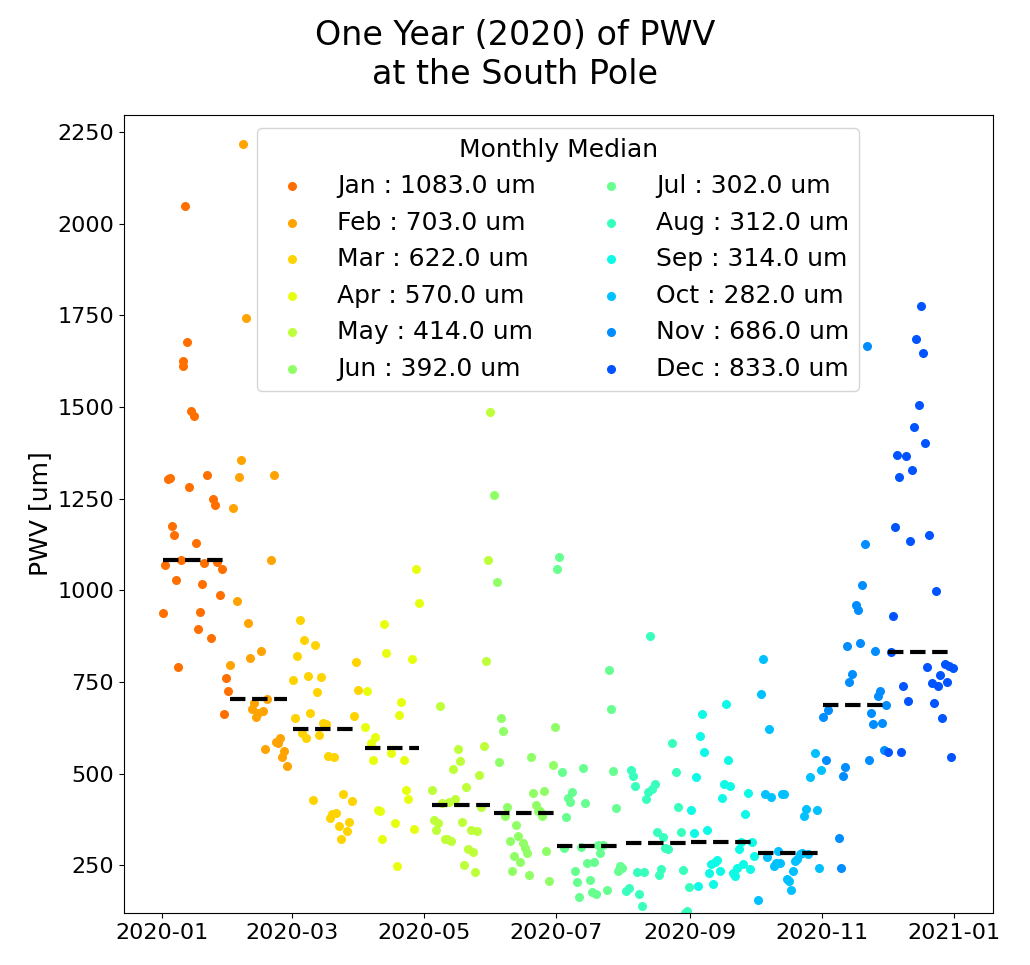} 
    \caption{Scatter plot showing PWV variations across different months of the year 2020. Monthly medians are indicated by dashed black lines and reported in the legend. }
    \label{fig:pwv_vs_month}
\end{figure}

Figure \ref{fig:pwv_vs_month} shows that the median PWV decreases steadily until July, remains relatively stable during July, August, and September, and reaches its annual minimum in October.
An analysis of zenith temperature fluctuations over the course of a year shows that PWV variations exhibit similar amplitudes on weekly and monthly timescales. The typical monthly PWV standard deviation is approximately $200$ $\mu$m.
On shorter timescales, PWV remains stable for periods of less than approximately 5 minutes, with a standard deviation of around $10$ $\mu$m over 1-minute intervals, as derived from sky dip observations.

\begin{figure}[htb!]
   \includegraphics[width=1.\columnwidth]{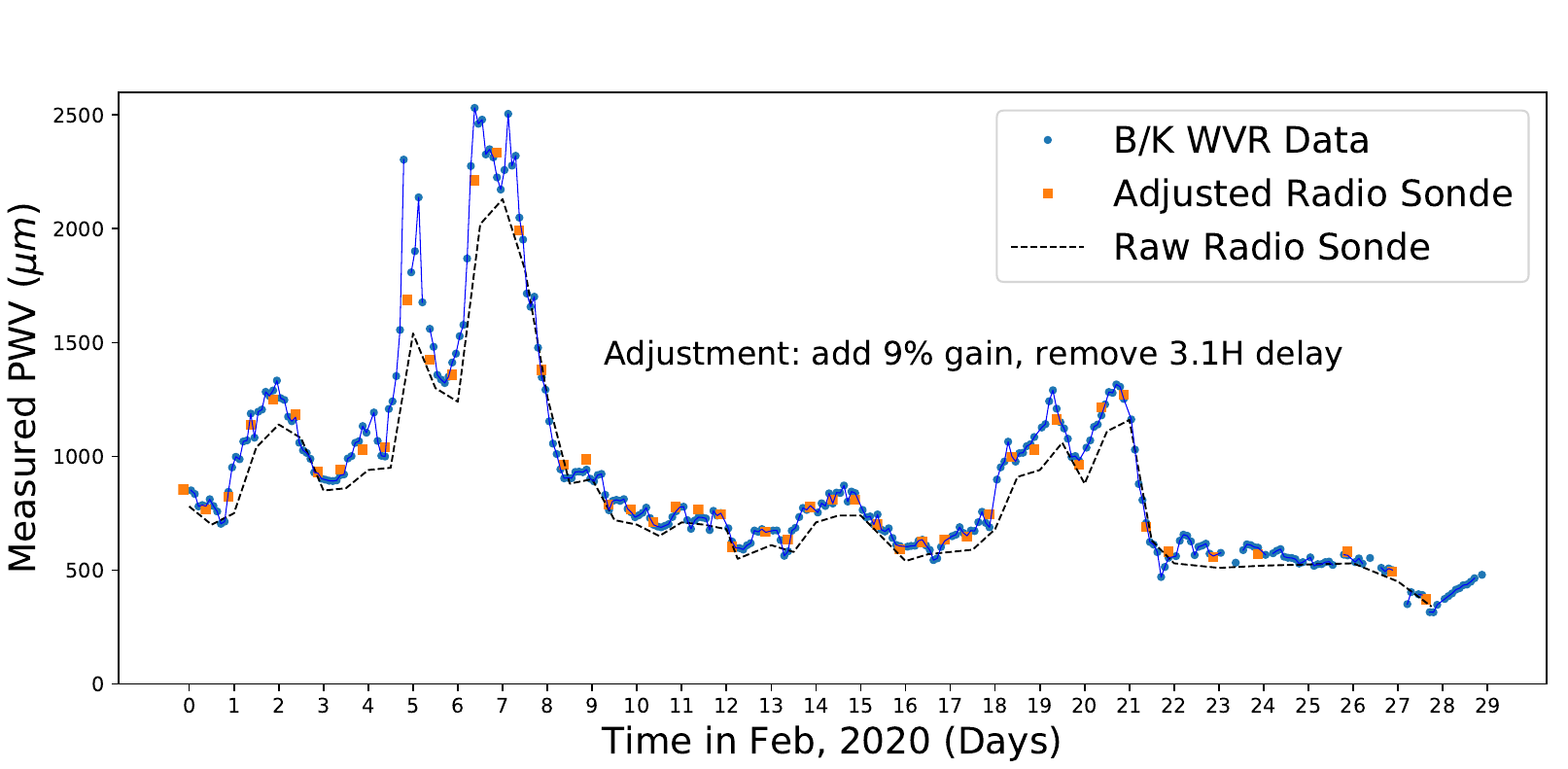} 
    \caption{Comparison between PWV values extracted from radiosonde launches and WVR observations acquired in February 2020. Each radiosonde data point corresponds to a single radiosonde launch, while each WVR data point corresponds to a single sky dip measurement. To account for the delay between the balloon launch and the availability of the processed radiosonde data, the radiosonde timestamps were shifted by the appropriate time offset. A 9\% gain correction was then applied to the radiosonde PWV values, after which the two datasets show excellent agreement. While the radiosonde provides one PWV measurement every 12 hours, the WVR delivers much higher temporal resolution and enables continuous mapping of PWV in both azimuth and elevation.}
    \label{fig:wvr_vs_rs}
\end{figure}

Figure \ref{fig:wvr_vs_rs} shows the excellent agreement between the PWV values extracted from WVR data and those independently measured by radiosonde. The WVR dataset consists of one month of zenith PWV measurements obtained from direct zenith observations via sky dips during the austral summer of 2020. 
During the same period, radiosonde data were acquired twice daily \citep{noaa}. The radiosonde data exhibit a delay of approximately 3 hours relative to the WVR measurements, due to the time lag between data acquisition and upload. After correcting for this delay, the WVR and radiosonde PWV estimates agree within $10\%$, confirming both the proper functioning of the WVR and the accuracy of the PWV extraction method. This comparison also highlights the superior temporal resolution of the WVR data.

 By utilizing direct zenith temperature measurements, we achieved high time resolution, which could be further improved, up to about $20$ Hz, through extrapolation from mid-elevation measurements. 

\subsection{PWV Atmograms and Wind Speed}
\label{subsec:pwv_fluctuations_Az}

\begin{figure}[htb!]
    \centering

    \annotatedimage{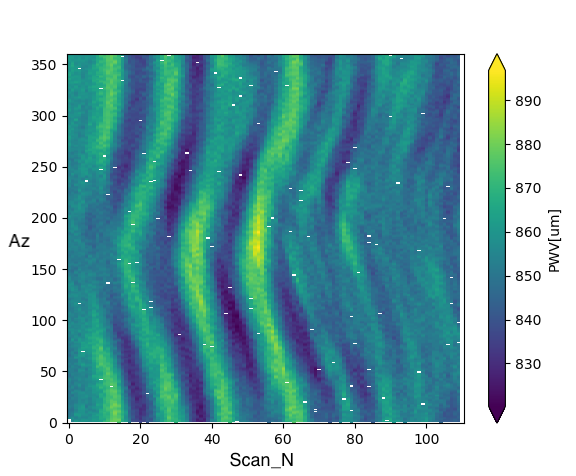}{(a)}

    \vspace{-4em}

    \annotatedimage{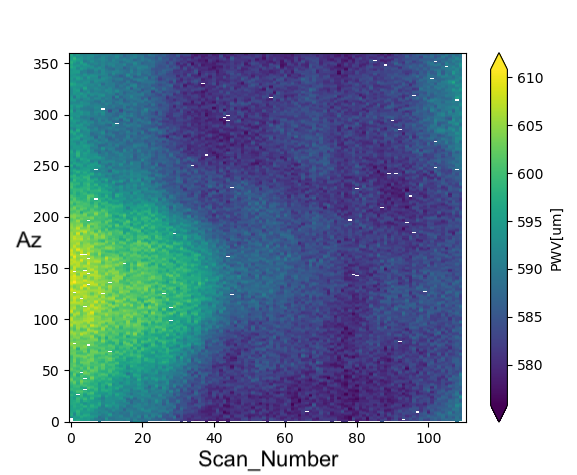}{(b)}

    \vspace{-4em}

    \annotatedimage{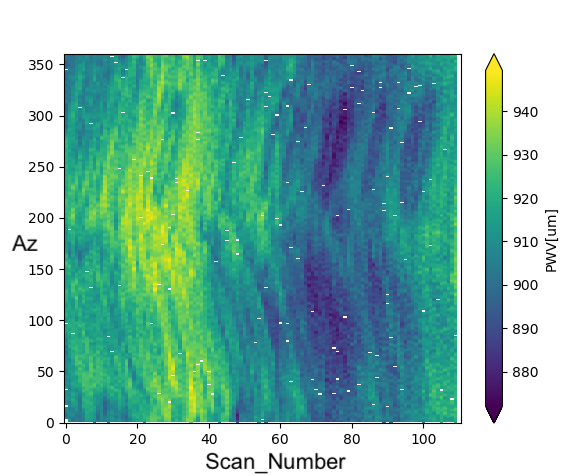}{(c)}

   \caption{Three representative PWV atmograms illustrating different atmospheric conditions: (a) 01/25/2020, showing coherent spatial structures consistent with a frozen atmospheric slab; (b) 01/17/2020, dominated by temporal variability with little coherent spatial structure; and (c) 02/05/2020, showing a combination of coherent spatial structure and temporal variability.}
    \label{fig:PWV_atmo_wind}
\end{figure}

In addition to providing average PWV estimates, the WVR also enables mapping of spatial PWV fluctuations.
We present PWV space-time fluctuations in the form of atmograms.\newline
An atmogram is a time series of successive azimuth scans, forming a two-dimensional map with azimuth along the y-axis and time along the x-axis. \newline 
Figure \ref{fig:PWV_atmo_wind} presents three representative PWV atmograms, which qualitatively display different atmospheric conditions. Each dataset spans 55 minutes and was acquired on different days, highlighting the variability in atmospheric behavior.
Atmogram (a) shows PWV fluctuations that behave like  spatial structures frozen in a slab of atmosphere moving rigidly at a constant wind speed.
Atmogram (b) is dominated by temporal fluctuations and lacks well-defined spatial structures.
Atmogram (c) is a combination of the behaviors seen in (a) and (b), showing fluctuations consistent with spatial structures frozen in a slab moving at a constant wind speed, on a background of temporal variability.
The distinct V-shaped pattern observed in atmograms (a) and (c) is produced by PWV structures moving across the sky and entering or exiting the WVR's field of view while the instrument is scanning.
When atmospheric PWV fluctuations can be approximated as a coherent, frozen slab advected across the sky by a constant wind speed and direction, atmograms can serve as an effective tool for extracting wind angular speed and direction.
The basic idea is that the time delay between timestream pairs at different azimuth (Az) positions, separated by a given $\Delta Az$, can be used to constrain the wind's angular speed and direction. 
If the atmospheric slab containing the PWV structures moves as a rigid body, the corresponding timestreams will exhibit similar signals, offset by a time delay that reflects the wind speed. 
To quantify this, we computed the cross-correlation between all possible pairs of timestreams and extracted the time delay ($\Delta t$) from the peak of the resulting correlation function.

Finding the peak of the correlation function as a function of azimuth, 
we obtain a sinusoidal signal whose amplitude is directly related to the angular wind speed $\omega_{wind}$, according to:
\begin{equation}
\omega_{wind}=\text{tan}(el)\cdot \frac{\Delta Az}{\Delta t} \text{sin}(\Delta Az - \phi_{wind}),
\end{equation}
where $\Delta t$ is the time delay between timestreams separated by $\Delta Az$, $\phi_{wind}$ is the wind direction relative to the scan reference frame and $el$ is the elevation the radiometer is pointing at during the scan ($el=55 \degree$).

\begin{figure}[htb!]
    \centering
    \subfigure[]{\includegraphics[width=\columnwidth]{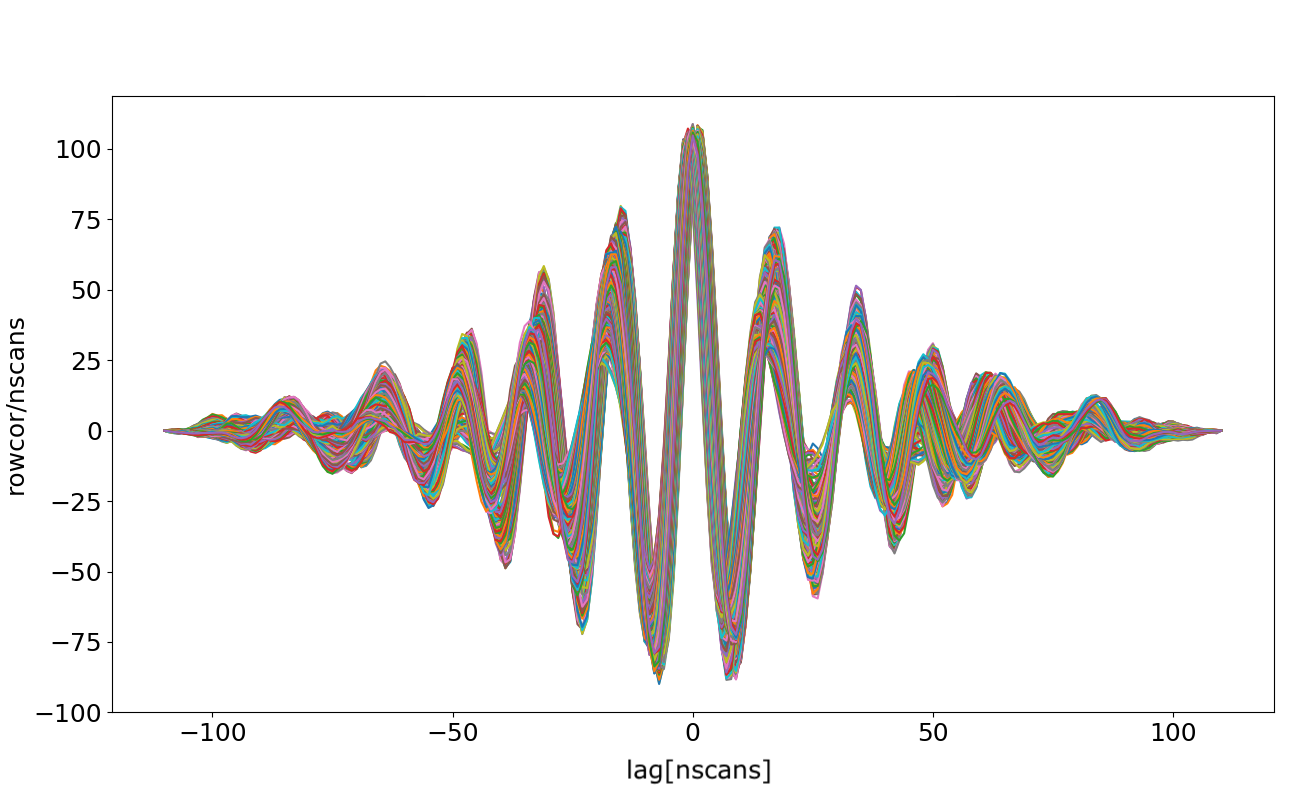}}
    \subfigure[]{\includegraphics[width=\columnwidth]{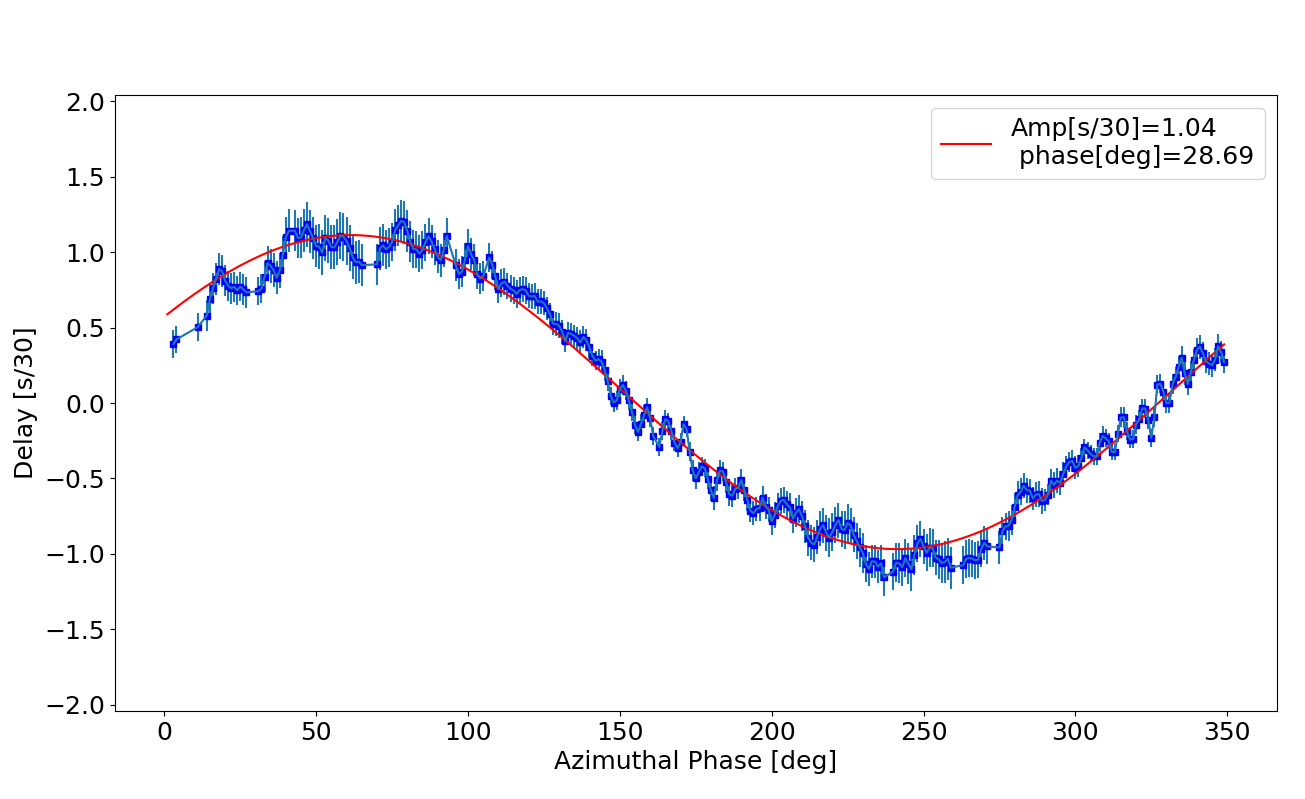}}
    \caption{Wind speed calculation from the WVR atmogram (a) in Fig.~\ref{fig:PWV_atmo_wind}. (a) The correlation function for each pair of timestreams with an angular separation of $\Delta \text{Az} = 14\degree$. (b) The corresponding delay plotted as a function of the azimuth phase. The y-axis is given in units of s/30, since one full azimuth scan corresponds to 30 s of data.}
    \label{fig:sky_real_delay}
\end{figure}

By comparing the angular wind speed derived from atmograms with the linear wind speed data provided by NOAA \citep{noaa} for the corresponding time period, we can estimate the effective height $h$ of the observed PWV fluctuations using the relation $v_{wind} = h \cdot \omega_{wind}$.
The results of this analysis, applied to atmogram (a) in Fig.~\ref{fig:PWV_atmo_wind}, are presented in Fig.~\ref{fig:sky_real_delay}.
From this comparison, we inferred an effective height of approximately 1.2 km, which falls near the center of the distribution reported by \cite{Bussmann_2005}.
Overall, our analysis suggests that the effective height of PWV structures varies from day to day, typically ranging between 0.5 km and 2 km.

\section{Atmospheric Noise in CMB Data}
\label{sec: atmo_noise_leak}

Emission from atmospheric water vapor, which is unpolarized [\cite{Errard_2015}, \cite{domec-atmo}], introduces variable foreground noise into CMB temperature measurements. 
Each Keck focal-plane pixel consists of two orthogonally polarized detectors that observe the same point on the sky. The detector pair sum provides a measurement of the total intensity (temperature), while the pair difference yields a measurement of the sky polarization, and removes the common-mode atmospheric emission. Because a single detector pair measures only one linear polarization orientation, observations at multiple boresight rotation angles are combined to reconstruct the Stokes Q and U polarization of the sky.
Instrumental systematics, such as bandpass mismatch or relative gain mismatch between the two orthogonally polarized detectors within a pixel, can then convert the water vapor unpolarized emission into spurious signals that leak into the CMB polarization data.
In this section, we present the results on the correlation between data acquired by the WVR and data obtained from the Keck high-frequency receiver at 270 GHz, a band which is particularly susceptible to atmospheric fluctuations. The same analysis was done on Keck 210 GHz data, yielding similar results. The plots presented here for the 270 GHz data are representative of both cases.
To facilitate comparison, the BICEP data are reported in the same atmogram format used for PWV maps in previous sections, and the WVR atmograms were converted from PWV units to temperature ($T_{\text{cmb}}$) at 270 GHz.
To convert PWV to temperature, we first extracted a full Rayleigh-Jeans temperature spectrum from PWV, using \texttt{am}, for each data point. The spectrum was then integrated over the Keck 270 GHz passband to compute the corresponding $T_{\text{RJ}}$ value. This process was repeated for each point in the atmogram, covering all azimuth positions and scan numbers. This results in a $T_{\text{RJ}}$ atmogram at 270 GHz.

\begin{figure}[htb!]
    \centering
    \includegraphics[width=\columnwidth]{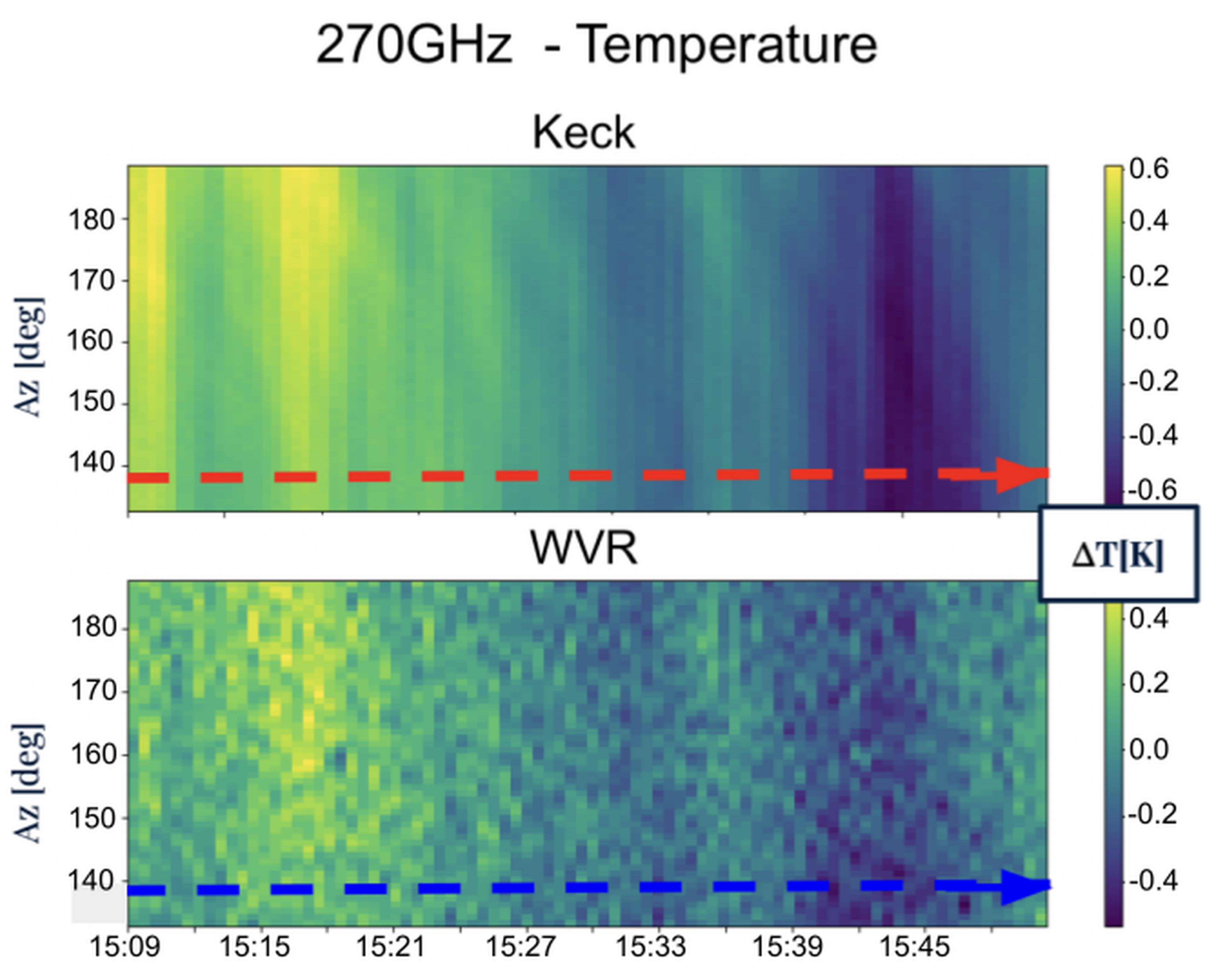}
    \caption{Temperature data from a single pixel of the Keck 270 GHz receiver (top) alongside the corresponding WVR data (bottom), which have been converted to equivalent temperature units in the Keck 270 GHz passband. The two arrows indicate the locations and directions of the data slices taken to enable direct timestream comparisons. This segmentation facilitates correlation analyses at fixed azimuth positions over extended timescales, mitigating the impact of WVR noise.}
    \label{fig:Az_fixed_cut_scheme}
\end{figure}

Subsequently, these Rayleigh-Jeans atmograms were converted into CMB-Planck temperature units. \newline
Atmograms from both instruments were aligned to correspond to identical spatial points and overlapping time ranges. 
All the analysis presented here was done on a per-pixel basis. The selected pixels are those observing regions of the sky located within one WVR beam from an elevation of $55\degree$, which corresponds to both the target elevation of the WVR and the center of the BICEP maps.
The correlation between the WVR atmograms and the atmograms extracted from Keck data is visually evident from the atmograms. \newline 
Since the WVR is a warm radiometer, unlike superconducting cryogenic detectors, it is inherently noisier. As a result, a direct one-to-one correlation between WVR azimuth scans\footnote{A Keck scan is defined as a complete back-and-forth movement of the mount across the full azimuth range. A half-scan covers this range in a single direction only.} and Keck half-scans at comparable sensitivity is not possible.
Each WVR azimuth scan lasts approximately 30 s, a timescale over which atmospheric variations are too small to produce a strong signal in the WVR data.

\begin{figure}[htb!]
    \centering

      \subfigure[]{\includegraphics[width=\columnwidth]{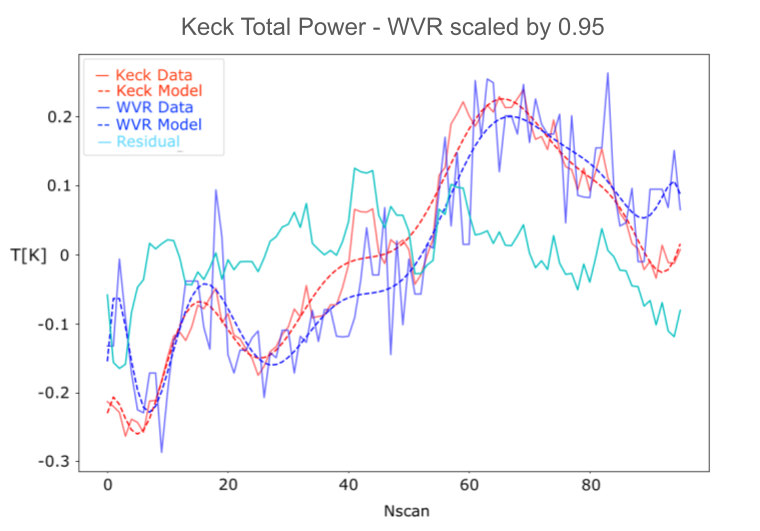}}
      \subfigure[]
     {\includegraphics[width=\columnwidth]{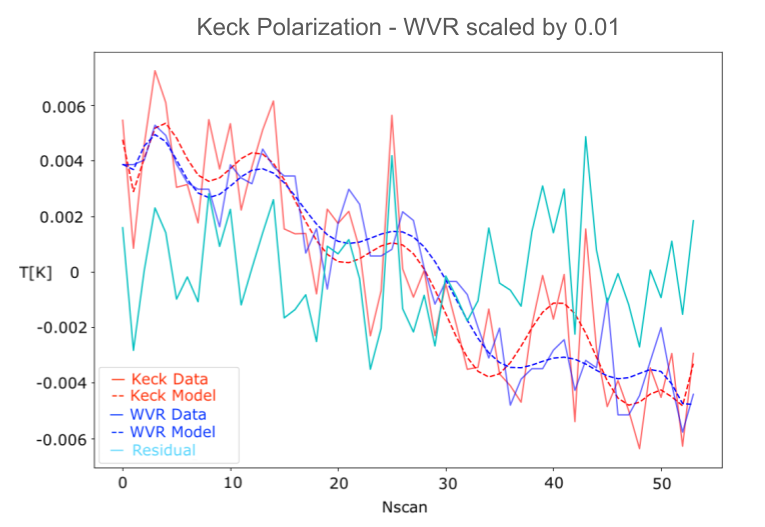}}

    \caption{Comparison between azimuth-fixed timestreams in $T_{\text{CMB}}$ units extracted from WVR data (blue) and the corresponding temperature and polarization timestreams measured by a Keck 270 GHz pixel (red). 
    The dashed blue line represents a 12th-order polynomial fit to the WVR data, used to suppress high-frequency noise and capture large-scale atmospheric trends. The cyan line shows the residual obtained by subtracting the rescaled polynomial model from the Keck data.
    The top panel shows a comparison between WVR and Keck temperature timestreams, while the bottom panel shows a comparison for polarization data.
    The WVR and Keck timestreams are drawn from different datasets to illustrate varying atmospheric conditions and structures at different observing times. }
    \label{fig:psum_pdiff_PWV_TOD}
\end{figure}
\begin{figure}[htb!]
    \centering
    \includegraphics[width=1\columnwidth]{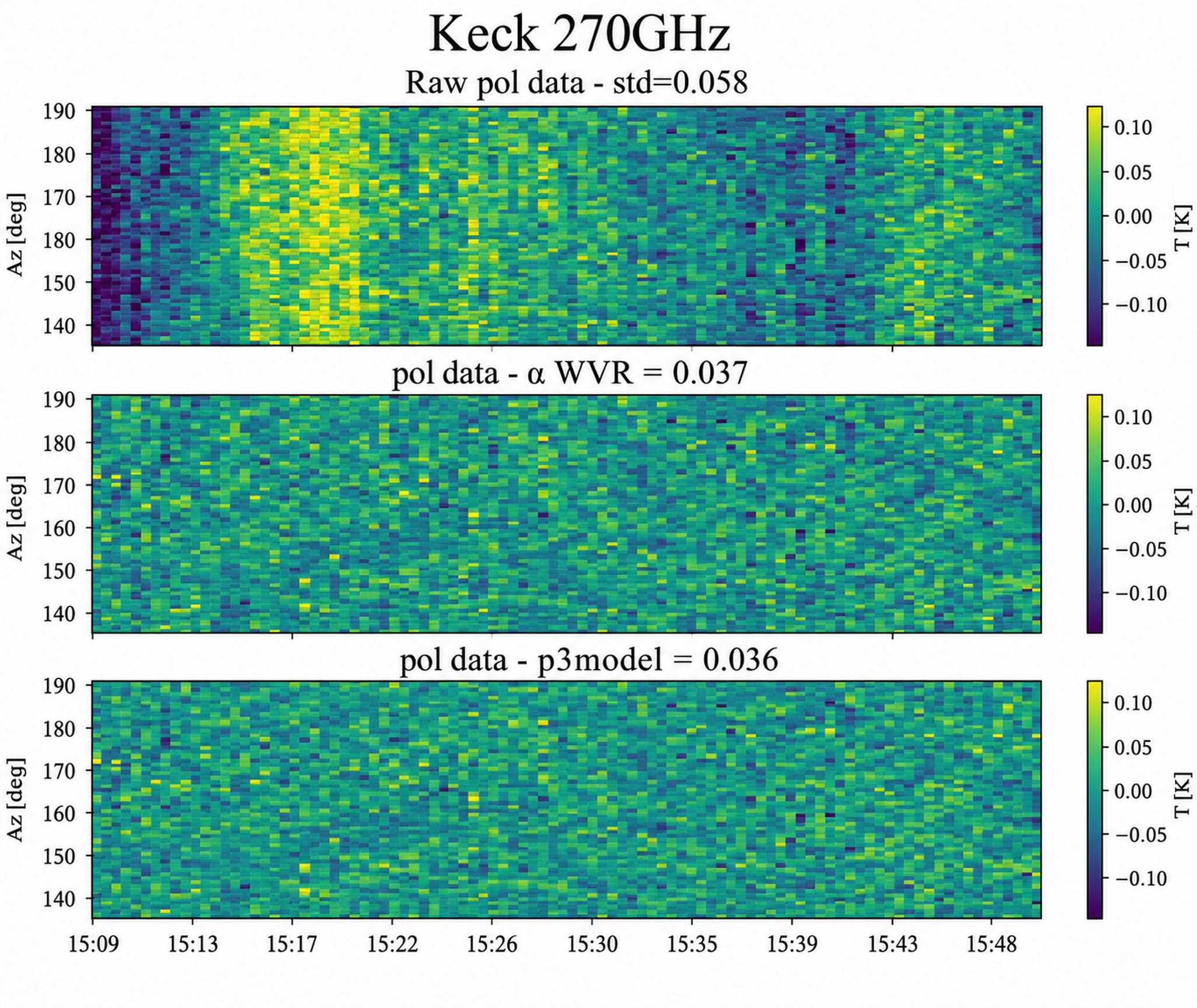}
    \caption{Comparison between the raw polarization data for a single pixel from the Keck 270 GHz receiver and the corresponding dataset after applying the WVR filtering described in Eq. \ref{eq:wvr_filt} and after subtracting a third order polynomial model to each half-scan. The final noise levels achieved using both methods for cleaning the scans are comparable.}
    \label{fig:atmo_wvrcleaned_pdiff_270}
\end{figure}

While one might consider correlating individual WVR azimuth scans with Keck half-scans, the low signal-to-noise ratio (SNR) in short WVR observations makes this approach ineffective. High SNR in the WVR requires integration over longer timescales, during which atmospheric fluctuations become more prominent.
Therefore, we adopted the orthogonal approach of analyzing timestreams at fixed azimuth positions, over timescales of 30 minutes (Fig.~\ref{fig:Az_fixed_cut_scheme}).
Figure \ref{fig:psum_pdiff_PWV_TOD} shows the timestream comparison between WVR data and Keck 270 GHz measurements (temperature and polarization) for a single Keck pixel and a fixed azimuth direction as indicated in Figure \ref{fig:Az_fixed_cut_scheme}. This example is representative; other choices of Keck pixel and azimuth direction produce similar results.
Each panel shows the raw WVR and Keck data, a 12th-order polynomial fit to both datasets (dashed line), and the resulting residual.
The residual is computed by first determining a linear scale factor $\alpha$ that best aligns the two polynomial models, and then subtracting the scaled WVR model from the Keck timestream: 
\begin{equation}
Residual = Keck_{TOD} - \alpha \cdot  WVR_{p12model}.
\label{eq:wvr_filt}
\end{equation}
The scale factor $\alpha$ between the Keck temperature and WVR data is approximately 0.95, indicating that the majority of the signal observed in the Keck temperature timestream is attributable to atmospheric fluctuations captured by the WVR.
In contrast, the scale factor between the Keck polarization data and the WVR is $\alpha \sim 0.01$, suggesting that only about 1\% of the atmospheric fluctuations measured by the WVR leak into the CMB polarization data \citep{BKXI2019}.\newline
Figure \ref{fig:atmo_wvrcleaned_pdiff_270} shows a full atmogram of polarization data from a single 270 GHz Keck pixel, along with the results of two atmospheric cleaning methods applied to each half-scan: the WVR-based filtering described in Eq.~\ref{eq:wvr_filt}, and simple subtraction of a third-order polynomial fit.
The strong correlation between WVR and Keck data demonstrates that the signal measured by the WVR is a reliable tracer of the atmospheric emission contaminating the CMB timestreams. This result provides important validation for the use of WVR to characterize atmospheric conditions and supports their application in comparing observing sites for millimeter-wave observations.


\section{Conclusion} 
The quest to constrain inflation through observations of CMB polarization remains one of the foremost goals of modern cosmology, and the South Pole continues to provide one of the most favorable sites on Earth for such observations. Despite the exceptionally stable atmospheric conditions at this location, atmospheric fluctuations persist as a significant source of noise in BICEP data.\newline
The analysis presented in this work demonstrates that the WVR is an effective instrument for atmospheric characterization and that its data correlate strongly with CMB observations from the Keck high-frequency receivers. These correlations support the use of WVR to characterize atmospheric conditions and to compare observing sites for millimeter-wave astronomy. In addition, they provide independent validation of the atmospheric subtraction methods used in BICEP analyses and improve our understanding of the removed signal.\newline
The WVR data show strong potential for atmospheric modeling in the context of CMB observations, but several limitations remain. Because the WVR is not a cryogenic detector, its intrinsic noise prevents reliable correlations with Keck timestreams on timescales shorter than a few minutes. To mitigate this, our analysis focuses on slowly sampled timestreams over longer integrations. In addition, the WVR operates independently of the BICEP telescope, scanning the sky in azimuth at a constant elevation with its own cadence, and it is insensitive to polarization. Although atmospheric emission is largely unpolarized, small polarized components may still be present \citep{Takakura2018,Fujino2024,Petroff2020}.\newline
Future implementations with lower-noise, polarization-sensitive radiometers co-aligned with the main receivers could further improve atmospheric noise mitigation and enhance the fidelity of CMB polarization measurements. In particular, a radiometer capable of measuring both total intensity and polarization would provide valuable information about the polarization properties of the atmosphere itself, an aspect that remains poorly characterized. \newline
Moreover, expanding the radiometer frequency coverage to include additional atmospheric lines, such as the 118 GHz oxygen transition, would enable simultaneous monitoring of multiple atmospheric constituents within or near the CMB observing bands and help disentangle the spectral origin of atmospheric contamination.\newline
These developments would lead to a more complete physical model of the atmosphere and provide new tools to better understand and mitigate systematic effects, such as temperature-to-polarization leakage, that currently limit the precision of CMB polarization measurements.

\section*{Acknowledgments}

The BICEP/Keck experiments have been funded through U.S. National 
Science Foundation grants, most recently including 2220444--2220448, 
2216223, 1836010, and 1726917. The development and deployment of the 
South Pole water vapor radiometer were supported by the Kavli Institute 
for Cosmological Physics at the University of Chicago, the 
Harvard-Smithsonian Center for Astrophysics, and Harvard University.

The research was carried out in part at the Jet Propulsion Laboratory, 
California Institute of Technology, under a contract with the National 
Aeronautics and Space Administration (80NM0018D0004). Focal plane 
development and testing were supported by the Gordon and Betty Moore 
Foundation at the California Institute of Technology. Readout 
electronics were supported by the Canada Foundation for Innovation 
through a grant to the University of British Columbia. The analysis 
effort at Stanford University and the SLAC National Accelerator 
Laboratory was partially supported by the U.S. Department of Energy. 
The computations in this paper were run on the Cannon cluster supported 
by the FAS Division of Science Research Computing Group at Harvard 
University.

We thank the staff of the U.S. Antarctic Program and, in particular, 
the staff of Amundsen-Scott South Pole Station, without whose support 
this research would not have been possible. We also thank our 
winter-over operators: Manwei Chan, Karsten Look, Calvin Tsai, 
Paula Crock, Ta Lee Shue, Grantland Hall, Hans Boenish, Robert Schwarz, 
Sam Harrison, Anthony DeCicco, Thomas Leps, Brandon Amat, Nathan Precup, 
Steffen Richter, Thibault Romand, Danielle Simmons, Markus Ayasse, 
Steven Jungst, Nathan McReynolds, and John Della Costa.

\bibliographystyle{aasjournal}
\bibliography{biblio}

@techreport{alma1-Emrich2009,
  author      = {Emrich, A. and Johansson, M. and Thidevall, R.},
  title       = {The {ALMA} Water Vapor Radiometer},
  institution = {Atacama Large Millimeter/submillimeter Array},
  number      = {ALMA Memo 587},
  year        = {2009},
}

@article{alma2-Nikolic2013,
  author  = {Nikolic, B. and Hills, R. E. and Richer, J. S. and Radford, S. J. E. and Stirling, A.},
  title   = {The {ALMA} Water Vapour Radiometer System},
  journal = {Astronomy \& Astrophysics},
  volume  = {552},
  eid     = {A104},
  year    = {2013},
  doi     = {10.1051/0004-6361/201220848}
}

@article{domec-atmo,
  author  = {Battistelli, E. S. and Amico, G. and Ba{\`u}, A. and Berg{\'e}, L. and Br{\'e}elle, {\'E}. and Charlassier, R. and Collin, S. and Cruciani, A. and de Bernardis, P. and Dufour, C. and Dumoulin, L. and Gervasi, M. and Giard, M. and Giordano, C. and Giraud-H{\'e}raud, Y. and Guglielmi, L. and Hamilton, J.-C. and Land{\'e}, J. and Maffei, B. and Maiello, M. and Marnieros, S. and Masi, S. and Passerini, A. and Piacentini, F. and Piat, M. and Piccirillo, L. and Pisano, G. and Polenta, G. and Rosset, C. and Salatino, M. and Schillaci, A. and Sordini, R. and Spinelli, S. and Tartari, A. and Zannoni, M.},
  title   = {Intensity and Polarization of the Atmospheric Emission at Millimetric Wavelengths at {Dome Concordia}},
  journal = {Monthly Notices of the Royal Astronomical Society},
  volume  = {423},
  number  = {2},
  pages   = {1293--1299},
  year    = {2012},
  doi     = {10.1111/j.1365-2966.2012.20951.x}
}

@article{Errard_2015,
  author  = {Errard, J. and others},
  title   = {Modeling Atmospheric Emission for {CMB} Ground-based Observations},
  journal = {The Astrophysical Journal},
  volume  = {809},
  number  = {1},
  eid     = {63},
  year    = {2015},
  doi     = {10.1088/0004-637X/809/1/63}
}

@dataset{noaa,
  author    = {Durre, Imke and Yin, Xungang and Vose, Russell S. and Applequist, Scott and Arnfield, Jeff and Korzeniewski, Bryant and Hundermark, Bruce},
  title     = {Integrated Global Radiosonde Archive ({IGRA}), Version 2},
  publisher = {NOAA National Centers for Environmental Information},
  year      = {2016},
  doi       = {10.7289/V5X63K0Q},
  note      = {Subset accessed for the 2020 South Pole analysis}
}

@dataset{merra,
  author    = {{Global Modeling and Assimilation Office}},
  title     = {{MERRA-2} Atmospheric Reanalysis Data},
  publisher = {NASA Goddard Earth Sciences Data and Information Services Center},
  year      = {2015},
  doi       = {10.5067/VJAFPLI1CSIV}
}

@manual{am_paine,
  author    = {Paine, Scott},
  title     = {The {am} Atmospheric Model},
  edition   = {Version 14.0},
  publisher = {Zenodo},
  year      = {2024},
  doi       = {10.5281/zenodo.13748391}
}

@article{Al_resistivity,
  author  = {Desai, P. D. and James, H. M. and Ho, C. Y.},
  title   = {Electrical Resistivity of Aluminum and Manganese},
  journal = {Journal of Physical and Chemical Reference Data},
  volume  = {13},
  number  = {4},
  pages   = {1131--1172},
  year    = {1984},
  doi     = {10.1063/1.555725}
}

@article{Bussmann_2005,
  author  = {Bussmann, R. S. and Holzapfel, W. L. and Kuo, C. L.},
  title   = {Millimeter Wavelength Brightness Fluctuations of the Atmosphere above the South Pole},
  journal = {The Astrophysical Journal},
  volume  = {622},
  number  = {2},
  pages   = {1343--1355},
  year    = {2005},
  doi     = {10.1086/427935}
}

@article{Guth_InflationaryUniverse,
  author  = {Guth, Alan H.},
  title   = {Inflationary Universe: A Possible Solution to the Horizon and Flatness Problems},
  journal = {Physical Review D},
  volume  = {23},
  number  = {2},
  pages   = {347--356},
  year    = {1981},
  doi     = {10.1103/PhysRevD.23.347}
}

@article{BK18,
  author  = {{BICEP/Keck Collaboration}},
  title   = {Improved Constraints on Primordial Gravitational Waves Using {Planck}, {WMAP}, and {BICEP/Keck} Observations through the 2018 Observing Season},
  journal = {Physical Review Letters},
  volume  = {127},
  number  = {15},
  eid     = {151301},
  year    = {2021},
  doi     = {10.1103/PhysRevLett.127.151301}
}

@article{BKXI2019,
  author  = {{BICEP2/Keck Array Collaborations}},
  title   = {{BICEP2/Keck Array XI}: Beam Characterization and Temperature-to-Polarization Leakage in the {BK15} Data Set},
  journal = {The Astrophysical Journal},
  volume  = {884},
  number  = {2},
  eid     = {114},
  year    = {2019},
  doi     = {10.3847/1538-4357/ab391d}
}

@inproceedings{Hui_BicepArrayPolarimeter,
  author    = {Hui, Howard and others},
  title     = {{BICEP Array}: A Multi-frequency Degree-scale {CMB} Polarimeter},
  booktitle = {Millimeter, Submillimeter, and Far-Infrared Detectors and Instrumentation for Astronomy IX},
  series    = {Proceedings of SPIE},
  volume    = {10708},
  eid       = {1070807},
  year      = {2018},
  doi       = {10.1117/12.2311725}
}

@article{barkats2018,
  author        = {Barkats, D. and Bowens-Rubin, R. and Clay, W. H. and Culp, T. and Hills, R. and Kovac, J. M. and Larsen, N. A. and Paine, S. and Sheehy, C. D. and Vieregg, A. G.},
  title         = {High-Precision Scanning Water Vapor Radiometers for Cosmic Microwave Background Site Characterization and Comparison},
  journal       = {arXiv e-prints},
  year          = {2018},
  eid           = {arXiv:1808.01349},
  eprint        = {1808.01349},
  archiveprefix = {arXiv},
  primaryclass  = {astro-ph.IM}
}

@article{Morris2025,
  author  = {Morris, Thomas W. and Battistelli, Elia and Bustos, Ricardo and Choi, Steve K. and Duivenvoorden, Adriaan J. and Dunkley, Jo and D{\"u}nner, Rolando and Halpern, Mark and Guan, Yilun and van Marrewijk, Joshiwa and Mroczkowski, Tony and Naess, Sigurd and Niemack, Michael D. and Page, Lyman A. and Partridge, Bruce and Puddu, Roberto and Salatino, Maria and Sif{\'o}n, Crist{\'o}bal and Wang, Yuhan and Wollack, Edward J.},
  title   = {The Atacama Cosmology Telescope: Quantifying Atmospheric Emission above Cerro Toco},
  journal = {Physical Review D},
  volume  = {111},
  number  = {8},
  eid     = {082001},
  year    = {2025},
  doi     = {10.1103/PhysRevD.111.082001}
}

@inproceedings{mackey2026,
  author    = {Mackey, S. and Papen, Alexander and Barkats, Denis and Barron, Darcy and Birdwell, Ian and Fatigoni, Sofia and Kovac, John M. and Paine, Scott and Petroff, Matthew A. and Vieregg, Abigail},
  title     = {Scanning Water Vapor Radiometers for {CMB} Observatories in Chile and at the South Pole},
  booktitle = {United States National Committee of URSI National Radio Science Meeting},
  pages     = {334--335},
  year      = {2026},
}

@article{Takakura2018,
  author  = {Takakura, Satoru and Aguilar Fa{\'u}ndez, M. A. O. and Akiba, Y. and Arnold, K. and Baccigalupi, C. and Barron, D. and Beck, D. and Bianchini, F. and Boettger, D. and Borrill, J. and Cheung, K. and Chinone, Y. and Elleflot, T. and Errard, J. and Fabbian, G. and Feng, C. and Goeckner-Wald, N. and Hamada, T. and Hasegawa, M. and Hazumi, M. and Howe, L. and Kaneko, D. and Katayama, N. and Keating, B. and Keskitalo, R. and Kisner, T. and Krachmalnicoff, N. and Kusaka, A. and Lee, A. T. and Lowry, L. N. and Matsuda, F. T. and May, A. J. and Minami, Y. and Navaroli, M. and Nishino, H. and Piccirillo, L. and Poletti, D. and Puglisi, G. and Reichardt, C. L. and Segawa, Y. and Silva-Feaver, M. and Siritanasak, P. and Suzuki, A. and Tajima, O. and Takatori, S. and Tanabe, D. and Teply, G. P. and Tsai, C.},
  title   = {Measurements of Tropospheric Ice Clouds with a Ground-based {CMB} Polarization Experiment, {POLARBEAR}},
  journal = {The Astrophysical Journal},
  volume  = {870},
  number  = {2},
  eid     = {102},
  year    = {2019},
  doi     = {10.3847/1538-4357/aaf381}
}

@article{Petroff2020,
  author  = {Petroff, Matthew A. and Eimer, Joseph R. and Harrington, Kathleen and Ali, Aamir and Appel, John W. and Bennett, Charles L. and Brewer, Michael K. and Bustos, Ricardo and Chan, Manwei and Chuss, David T. and Cleary, Joseph and Denes Couto, Jullianna and Dahal, Sumit and D{\"u}nner, Rolando and Essinger-Hileman, Thomas and Flux{\'a} Rojas, Pedro and Gothe, Dominik and Iuliano, Jeffrey and Marriage, Tobias A. and Miller, Nathan J. and N{\'u}{\~n}ez, Carolina and Padilla, Ivan L. and Parker, Lucas and Reeves, Rodrigo and Rostem, Karwan and Nunes Valle, Deniz Augusto and Watts, Duncan J. and Weiland, Janet L. and Wollack, Edward J. and Xu, Zhilei},
  title   = {Two-year Cosmology Large Angular Scale Surveyor ({CLASS}) Observations: A First Detection of Atmospheric Circular Polarization at {Q} Band},
  journal = {The Astrophysical Journal},
  volume  = {889},
  number  = {2},
  eid     = {120},
  year    = {2020},
  doi     = {10.3847/1538-4357/ab64e2}
}

@article{Fujino2024,
  author  = {Fujino, Takuro and Takakura, Satoru and Arani, Shahed Shayan and Barron, Darcy and Baccigalupi, Carlo and Chinone, Yuji and Errard, Josquin and Fabbian, Giulio and Feng, Chang and Halverson, Nils W. and Hasegawa, Masaya and Hazumi, Masashi and Jeong, Oliver and Kaneko, Daisuke and Keating, Brian and Kusaka, Akito and Lee, Adrian and Matsumura, Tomotake and Piccirillo, Lucio and Reichardt, Christian L. and Sakaguri, Kana and Siritanasak, Praween and Yamada, Kyohei},
  title   = {A Measurement of Atmospheric Circular Polarization with {POLARBEAR}},
  journal = {The Astrophysical Journal},
  volume  = {981},
  number  = {1},
  eid     = {15},
  year    = {2025},
  doi     = {10.3847/1538-4357/ada89b}
}

\end{document}